\PassOptionsToPackage{table}{xcolor}

\documentclass[sigconf, nonacm]{acmart}
\AtBeginDocument{%
  }

\usepackage{acmart-taps}

\copyrightyear{2026}
\acmYear{2026}
\setcopyright{cc}
\setcctype{by-nc-nd}
\acmConference[CHI '26]{Proceedings of the 2026 CHI Conference on Human Factors in Computing Systems}{April 13--17, 2026}{Barcelona, Spain}
\acmBooktitle{Proceedings of the 2026 CHI Conference on Human Factors in Computing Systems (CHI '26), April 13--17, 2026, Barcelona, Spain}
\acmDOI{10.1145/3772318.3791386}
\acmISBN{979-8-4007-2278-3/2026/04}

\begin{document}

\title{Living Contracts: Beyond Document-Centric Interaction with Legal Agreements}

\author{Ziheng Huang}
\affiliation{
 \institution{University of Illinois Urbana-Champaign}
  \city{Urbana}
  \country{USA}}
\email{zihengh2@illinois.edu}

\author{Robin Kar}
\affiliation{
 \institution{University of Illinois Urbana-Champaign}
  \city{Urbana}
  \country{USA}}
\email{rkr@illinois.edu}

\author{Hari Sundaram}
\affiliation{
 \institution{University of Illinois Urbana-Champaign}
  \city{Urbana}
  \country{USA}}
\email{hs1@illinois.edu}

\author{Tal August}
\affiliation{
 \institution{University of Illinois Urbana-Champaign}
  \city{Urbana}
  \country{USA}}
\email{taugust@illinois.edu}

\begin{abstract}
User interaction with legal contracts has been limited to document reading, which is often complicated by complex, ambiguous legal language. We explore possible futures where contract interfaces go beyond single document interfaces to (1) educate users with legal rights not stated in the contract, (2) transform legal language into alternative representations to aid information tasks before, during, and after signing, and (3) proactively supply contractual information at relevant moments. We refer to these future interfaces collectively as Living Contracts. Using residential leases as a case study, we created three design probes representing different possible Living Contracts. A three-part qualitative study (N=18) revealed participants' barriers to interacting with contracts, including interpreting complex language, uncertainty about legal rights, and the pressure to sign quickly. Participants’ feedback on the probes highlighted how Living Contracts have the potential to address these challenges and open new design opportunities for human-contract interactions beyond document reading.

\end{abstract}



\maketitle


\section{Introduction}

Legal contracts govern the conditions under which people live and work, from housing and employment to insurance and digital service usage. 
While contracts are interfaces through which people navigate their contractual rights and obligations, legal scholars have argued that consumer-facing contracts have drifted from their foundational premise of mutual understanding and cooperation ~\cite{kar2019pseudo, haapio2021contracts, nousiainen2022legal}.
Instead, contracts such as leases can be written in complex legal language ~\cite{furth2017unexpected, korobkin2003bounded, so2023small} and increasingly contain unenforceable clauses and obfuscating language that mislead consumers about their rights ~\cite{furth2017unexpected, hoffman2022leases, prescott2024subjective, hicks1972contractual, furth2018harmful}.

Past HCI and legal work has focused on supporting contract reading through information design, such as redesigning layouts~\cite{passera2015beyond}, integrating diagrams~\cite{passera2012enhancing, passera2016exploring, passera2017diagrams}, or using comics~\cite{ketola2024comic, tabassum2018increasing}. 
Yet, a growing movement in the legal community argues that contracts should be engaged with before and after signing, rather than just during ~\cite{haapio2021contracts, nousiainen2022legal}. 
To illustrate this argument with an example: a lease could inform a renter's choice about where to live (i.e., before signing) and represent obligations and rights a renter can reference later (i.e., after signing) rather than be seen as a necessary hurdle to renting an apartment (i.e., during signing).

Transforming textual information beyond a document has recently become possible with advances in language models (LMs), which enable dynamically retrieving ~\cite{yue2024lawllm, zheng2025reasoning, ryu2023retrieval} or transforming text into alternative representations~\cite{servantez2023computable}. While current LM technology is not mature enough to safely realize this vision of alternative contractual representations due to the high-stakes nature of legal text \citep{nielsen2025law, barbara2024automatic, kang2024using}, what might the emerging capabilities of LMs mean for future contract interfaces?

In this work, we speculate on what user interfaces for contracts could look like that support engagement before, during, and after signing --- a concept we term Living Contracts.  
In contrast with prior work on document design~\cite{passera2015beyond}, we explore the possibility of contract interfaces as something more than a single document, instead as interactive interfaces that can: (1) provide relevant background knowledge not present in the original contract, (2) transform contractual information into alternative representations to support information tasks across the contract lifecycle, and (3) proactively surface and advocate for users' contractual rights and obligations. 
Using leasing contracts as a case study to explore the vision of Living Contracts, we created three design probes implemented as interactive web interfaces that reconceptualize the renting process: \textbf{LeaseCompare} foregrounds contractual clauses to inform early apartment search and comparison; \textbf{LeaseRead} augments legal text with comics, explorable scenarios, and relevant legal information; and \textbf{LeaseTrack} proactively surfaces contractual obligations and rights after signing. 
Our goal is not to evaluate usability, but to explore what contract interfaces could become.

We conducted a three-part study with 18 participants having a wide variety of contractual experiences, including as both landlords and tenants: (1) initial semi-structured interviews on participants’ challenges with contracts; (2) interaction with three design probes to explore opportunities and concerns of future contract interfaces; and (3) a design ideation session.
Our findings from the initial semi-structured interviews revealed that participants often felt they lacked awareness of their legal rights and engaged with contracts only at the point of signing, leading to pressure and discomfort when signing. 
While interacting with the probes, participants saw value in how the probes contextualized, restructured, and proactively provided contractual information.  
For example, LeaseRead educated unexpected legal knowledge and its representation of contracts as explorable scenarios and comics prompted participants to envision possible outcomes of signing.
In addition, proactively supplying contractual information could inform decisions both before signing (e.g., apartment search) and after signing (e.g., encountering unlawful requests).
At the same time, some participants saw risks of alternative contract interfaces, such as misrepresenting contractual information or a diminished sense of autonomy from proactive reminders.
In the design ideation session, participants extended the design concepts of Living Contracts beyond leases (e.g., insurance, surgical, and loan) and proposed additional designs, including multi-contract orchestration. We end by discussing the implications of Living Contracts for inspiring novel contract interfaces and considerations for implementing them. In this paper, we make the following contributions:

\begin{enumerate}

\item We propose going beyond document-centric interaction with legal agreements and offer a reconceptualization of contract interfaces as contextual, interactive, and proactive interfaces that support user interaction across the contract lifecycle, referred to as Living Contracts. 

\item Using leases as a case study, we explore possible Living Contracts through the design of three interactive probes that augment user interaction with leases before (LeaseCompare), during (LeaseRead), and after signing (LeaseTrack).

\item We report on a three-part study that surfaced barriers to contract engagement and explored alternative contract interfaces through the probes. Our findings highlight the opportunities and risks of Living Contracts in resolving reported barriers to contract engagement. 

\end{enumerate}



\begin{figure*}[!ht]
    \centering
    \includegraphics[width=0.82\textwidth]{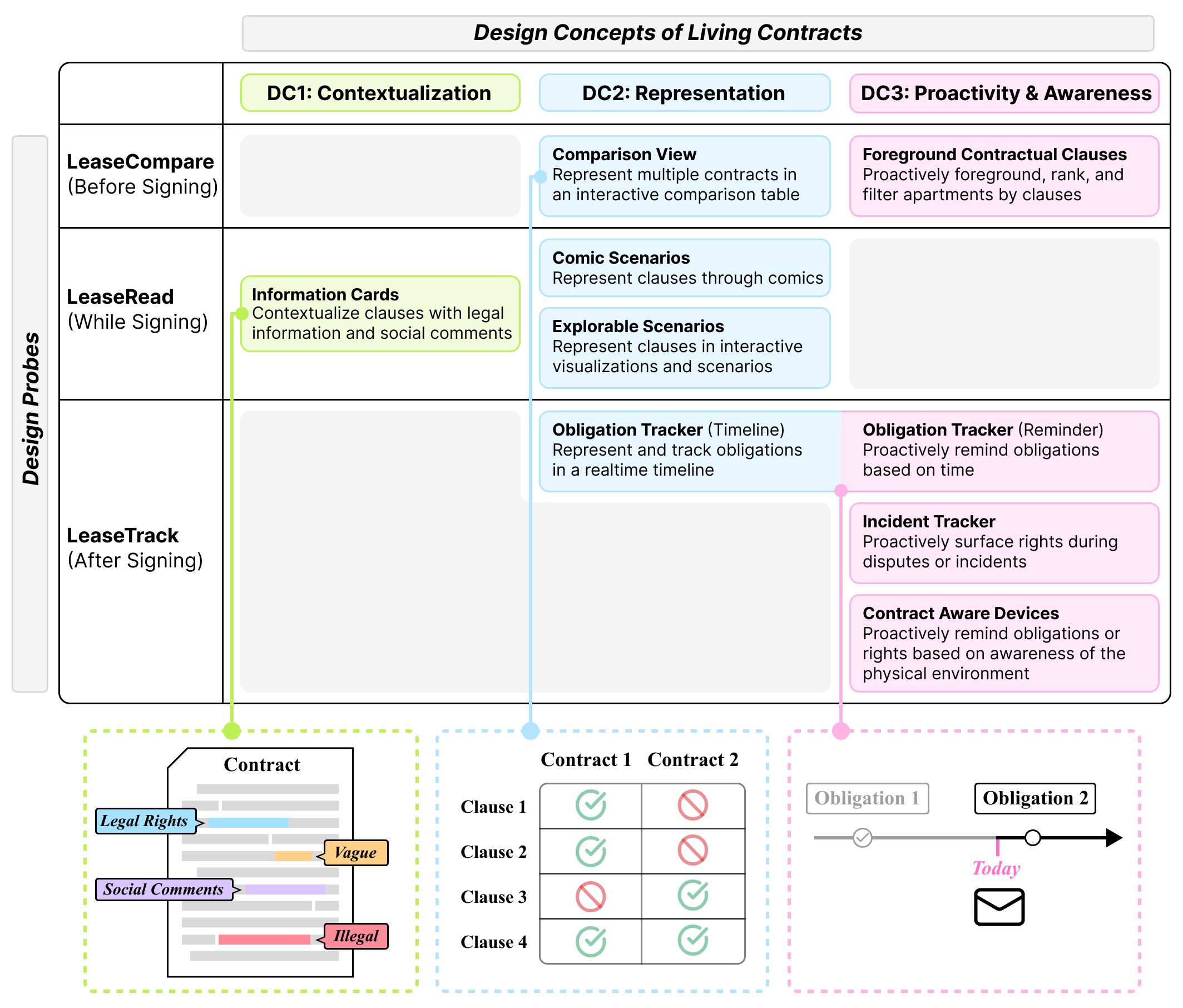}
    \caption{Overview of the three design probes (i.e., LeaseCompare, LeaseRead, LeaseTrack) and their features, mapped to the three design concepts of Living Contracts. Living Contracts is the idea that contract interfaces can go beyond static document readers and instead: contextualize contractual information with relevant background knowledge (DC1), transform contractual information into task-specific representations (DC2), and proactively surface and supply contractual information (DC3).}
    \label{fig:featuremapping}
    \Description{Overview of the three design probes and their features, mapped to the design concepts of Living Contracts. Features illustrating Contextualization: Information Cards that contextualize clauses with legal information and social comments. Features illustrating Representation: a comparison view for multiple contracts, comic scenarios, explorable scenarios, and an obligation-tracking timeline. Features illustrating Proactivity and Awareness: foregrounding contractual clauses during apartment search, proactively reminding obligations based on time, surfacing rights during disputes or incidents, and contract-aware devices that proactively remind users of obligations or rights based on the physical environment. At the bottom of the figure, three examples are illustrated: Comparison Table (e.g., restructure multiple contracts into a comparison table to aid apartment selection), Information Cards (e.g., contextualizing clauses with external legal knowledge to support contract reading), and Obligation Tracker (e.g., restructure obligations into a timeline tracker that proactively reminds users).}
\end{figure*}

\section{Background and Related Work}

\subsection{Consumer-Facing Contracts}

Much of the existing work in HCI has focused on digital contracts presented as clickwraps, such as privacy policies and Terms of Service (ToS), highlighting their length, complexity, and users’ widespread disengagement~\cite{obar2020biggest, bakos2014does}.
Yet, higher-stakes consumer-facing contracts such as leasing, loan, insurance, or employment contracts that carry direct and material consequences remain underexplored in HCI.
These high-stakes contracts can be written in extended length, complex language, and unfriendly layout that discourages consumer engagement, understanding, and obligation management~\cite{furth2017unexpected, mueller1970residential, so2023small}. 
More damaging, these contracts (e.g., leases, employment) often contain ambiguous language~\cite{furth2017unexpected, furth2018harmful} and unenforceable clauses~\cite{furth2017unexpected, hoffman2022leases, prescott2024subjective, hicks1972contractual} that obscure consumer rights or providers' responsibilities, making consumers less likely to seek information or advocate for themselves~\cite{furth2018harmful, prescott2024subjective}. 
Legal scholars have argued that modern consumer contracts have drifted away from their foundational premise of mutual understanding and cooperation. Instead, they function as one-sided instruments intended to minimize legal risk~\cite{kar2019pseudo, haapio2021contracts, nousiainen2022legal}. 

Legal scholars often describe contracts as unfolding across a lifecycle of stages including: (1) \textit{formation}, where parties bargain, negotiate, or decide whether to enter into an agreement ~\cite{furmston2010contract}; (2) \textit{interpretation}, where parties identify the rights and obligations that a contract implicates if formed ~\cite{mitchell2018interpretation}; (3) \textit{performance and breach}, where parties perform or breach the contract after signing~\cite{haapio2021contracts}; (4) \textit{defenses}, where a party argues that the contract should not be enforced after breaching (e.g., under duress, undue influence, or minority) ~\cite{smith1997contracting}; and (5) \textit{remedies}, where the law provides compensation or other relief in response to a breach ~\cite{farnsworth1970legal}.
HCI research has predominantly focused on contract reading during the interpretation phase (i.e., reading and signing the contract). 
Our study complements prior work by showing that challenges may arise not only from the contract text itself but also from the broader context in which contracts are read, signed, and enforced.
Through Living Contracts, we explore possible interfaces that cover interpretation (LeaseRead) while extending to formation (LeaseCompare) and performance (LeaseTrack) activities.

\subsection{Human-Centered Contracts}

While contracts are often seen as legal instruments for dispute resolution, legal scholars have argued that they should also be understood as interfaces to foster shared understanding, communication, and obligation fulfillment ~\cite{haapio2021contracts, nousiainen2022legal}. 
In response, the Legal Design and HCI communities have proposed recommendations to make contracts more user-friendly. 
\citet{kay2010textured} introduced Textured Agreements, which employ typographic cues, pull quotes, vignettes, and symbols to enhance attention and comprehension of privacy policies compared to plain text. Similarly, \citet{passera2015beyond} showed that visually redesigned information organization of tenancy agreements improved users' accuracy and efficiency in understanding contractual obligations. Other work has incorporated diagrams, such as flowcharts and swarm visualizations, to clarify clauses in B2B contracts ~\cite{passera2012enhancing, passera2016exploring, passera2017diagrams}. 
To improve engagement, researchers have also experimented with providing summaries with comics as a medium for contract communication ~\cite{ketola2024comic, tabassum2018increasing}. 
Works focusing on online ToS or privacy policies have also explored providing summaries to help consumers gain an overview of the policy ~\cite{good2005stopping, good2007noticing, kelley2009nutrition}. 
Complementing these efforts, Haapio and colleagues aggregated and curated design patterns and recommendations for contracts, including content organization (e.g., layered layouts) and visual representations (e.g., comics)~\cite{haapio2022visualisation, haapio2016design, haapio2017contracts}.
While prior work mainly focused on the reading process, our work focuses on exploring how contracts can be reimagined as interactive interfaces that support user engagement before, during, and after contract formation.

\subsection{Interactive Documents}

The HCI community has long explored how interactivity can enhance document reading.
Prior work has explored providing reading guidance for academic papers and news articles by highlighting important information~\cite{fok2024supporting, fok2022scim, chen2023marvista} or offering guiding questions~\cite{august2023paper}. For example, Scim highlights objectives, results, and methods in academic papers, to support skimming ~\cite{fok2024supporting, fok2022scim}.
Some systems focus on helping users clarify information through position-sensitive definitions~\cite{head2021augmenting}, plain language summaries~\cite{august2023paper, august2024know}, or allowing users to ask questions about the document~\cite{zhao2020talk, fok2024supporting}.
Other work has explored interactive interrogation of mathematical concepts or statistical reports. For example, Augmented Physics integrates interactive simulations into textbooks to help students explore physics concepts~\cite{gunturu2024augmented}. Math Augmentation and Augmented Math examines how interactivity and visual design can make mathematical formulas more accessible~\cite{head2022math, chulpongsatorn2023augmented}. Explorable multiverse analysis reports enable readers to interactively explore how different analytical choices affect research findings~\cite{dragicevic2019increasing}. In this work, we take inspiration from prior work to explore how interactivity might augment an underexplored category of document in HCI: legal contracts.

\subsection{Emerging Capabilities for AI Text Transformation}

Recent advances in AI have enabled new opportunities for automated retrieval and transformation of legal documents.
One thread of research focuses on automatically retrieving relevant legal information for legal research and question answering, such as prior cases or statutes~\cite{yue2024lawllm, zheng2025reasoning, ryu2023retrieval}. 
Another line of work seeks to automatically identify red flags in a contract ~\cite{leivaditi2020benchmark, hendrycks2021cuad}.
Recently, work has also explored transforming legal documents into summaries or machine-readable formats. For example, ~\citet{sancheti2023read} trained models to generate party-specific summarizations of Obligations, Entitlements, and Prohibitions of leasing contracts. 
~\citet{servantez2023computable} extracted key entities, relationships between entities, and formulas to form obligation logic graphs for representing the contractual obligations between parties. 
Additionally, ~\citet{barbara2024automatic} and ~\citet{kang2024using} attempted to automatically transform insurance contracts into computer code, highlighting the limitations of current LMs. 
While these technical advances hint at a future where legal documents can be flexibly retrieved and transformed, little is known about how to design the interfaces that leverage these capabilities to support users throughout the contract lifecycle. This work explores design opportunities for contract interfaces that could transform how people interact with legal documents.



\section{Methodology}

Speculative design has long been used in HCI for exploring alternative and future technologies. It intentionally suspends assumptions about what is likely or feasible to surface what is possible, desirable, or problematic ~\cite{dunne2024speculative, galloway2018speculative, bardzell2013critical}. 
To extend the speculation beyond designers themselves, researchers increasingly involve participants to co-speculate about possible futures through the use of probes. These probes can span a wide range of forms in illustrating alternative technologies, such as comic scenarios~\cite{wong2023broadening}, UI mockups~\cite{chow2025beyond}, or interactive interfaces~\cite{salovaara2025triangulating, noortman2019hawkeye}. 
Below, we describe our research process. We start by describing the design concepts that inspired Living Contracts (\S\ref{sec:designconcept}). 
Next, we introduce three design probes that illustrate Living Contracts in the context of leasing (\S\ref{sec:designprobes}), along with details about each probe (\S\ref{subsec:leasecompare}, \S\ref{subsec:leaseread}, \S\ref{subsec:leasetrack}). Lastly, we describe a three-part study we conducted using the probes (\S\ref{sec:method}).

\subsection{Living Contracts: Design Concepts}
\label{sec:designconcept}

Traditionally, signers’ interactions with contractual information have been limited to the written document. We explore future interfaces that go beyond document-centric interaction with contractual information. We refer to these interfaces as Living Contracts. Through four rounds of discussions with HCI and Legal scholars within the research team over a month, we synthesized three design concepts from prior literature in HCI and Law that inform Living Contracts (more details in Appendix~\ref{subsubsec:designprocess}).

\subsubsection{\textbf{DC1: Contextualization}}

In legal scholarship, ``law in the books'' refers to the formal laws and regulations, while ``law in the world'' captures how laws are applied and experienced in practice ~\cite{pound1910law}. While contracts are often presented as standalone documents for consumers to sign, interpreting a clause often requires both types of knowledge ~\cite{mitchell2018interpretation}. For example, a lease clause about late fees might seem valid in the contract to consumers, but could be unenforceable under local housing law, or enforced more leniently in practice. This motivated us to imagine contracts as agreements contextualized with relevant background knowledge.

\subsubsection{\textbf{DC2: Malleable Representation}}

While contracts are treated primarily as legal tools to minimize risk, legal scholars have advocated that they are also communication tools to foster mutual understanding ~\cite{kar2019pseudo, haapio2021contracts, nousiainen2022legal}.
The HCI community has proposed the idea of going beyond document-centered interactions by treating information as a dynamic and malleable entity, capable of being structured into different representations to support emerging tasks~\cite{fox2020towards}. 
Drawing on this perspective, we explore alternative representations of contracts that aid information tasks before, during, and after contract formation (e.g., signing).

\subsubsection{\textbf{DC3: Proactivity and Situation Awareness}}

Legal scholars have argued that contracts should function not only to resolve disputes after they arise but also to anticipate problems and promote successful collaboration beforehand, an approach termed `proactive' or `preventive' law ~\cite{haapio2021legal, haapio2021contracts}.
In this work, we consider how contract interfaces might be proactive: surfacing relevant information at appropriate moments and transitioning from a passive record into an active participant in signers' ongoing decisions.

\subsection{A Case Study on Leasing Contracts}
\label{sec:designprobes}


We used leasing contracts as a case study to create three design probes implemented as interactive web interfaces that illustrate the concepts of Living Contracts (\S\ref{sec:designconcept}). Figure ~\ref{fig:featuremapping} offers an overview of the probes mapped to each design concept.
Similar to the original intention of probes ~\cite{gaver1999design}, 
our goal with the probes was to elicit participants’ imagination and reflection on future contract interfaces rather than to test usability or derive conclusive design requirements ~\cite{gaver1999design, boehner2007hci, noortman2019hawkeye, chow2025beyond}.
We situated the three probes within a fictional company, Living Contract Corporation, to encourage participants to see them as interconnected services spanning the entire leasing process. The company offers three services that help tenants search for an apartment using contract information (LeaseCompare: \S\ref{subsec:leasecompare}), read a lease contract (LeaseRead: \S\ref{subsec:leaseread}), and manage their lease obligations after signing (LeaseTrack: S\ref{subsec:leasetrack}). In the sections below, we describe each of the three probes. 
All the legal information in the probes was validated through 3 one-hour consultation sessions with two contract lawyers with experience in leases. 
Additional details about our design process, probe implementation, and narrative scenarios to ground the probes can be found in Appendix~\ref{Appendix:additionaldetails}.

\begin{figure}[h]
    \centering
    \includegraphics[width=0.46\textwidth]{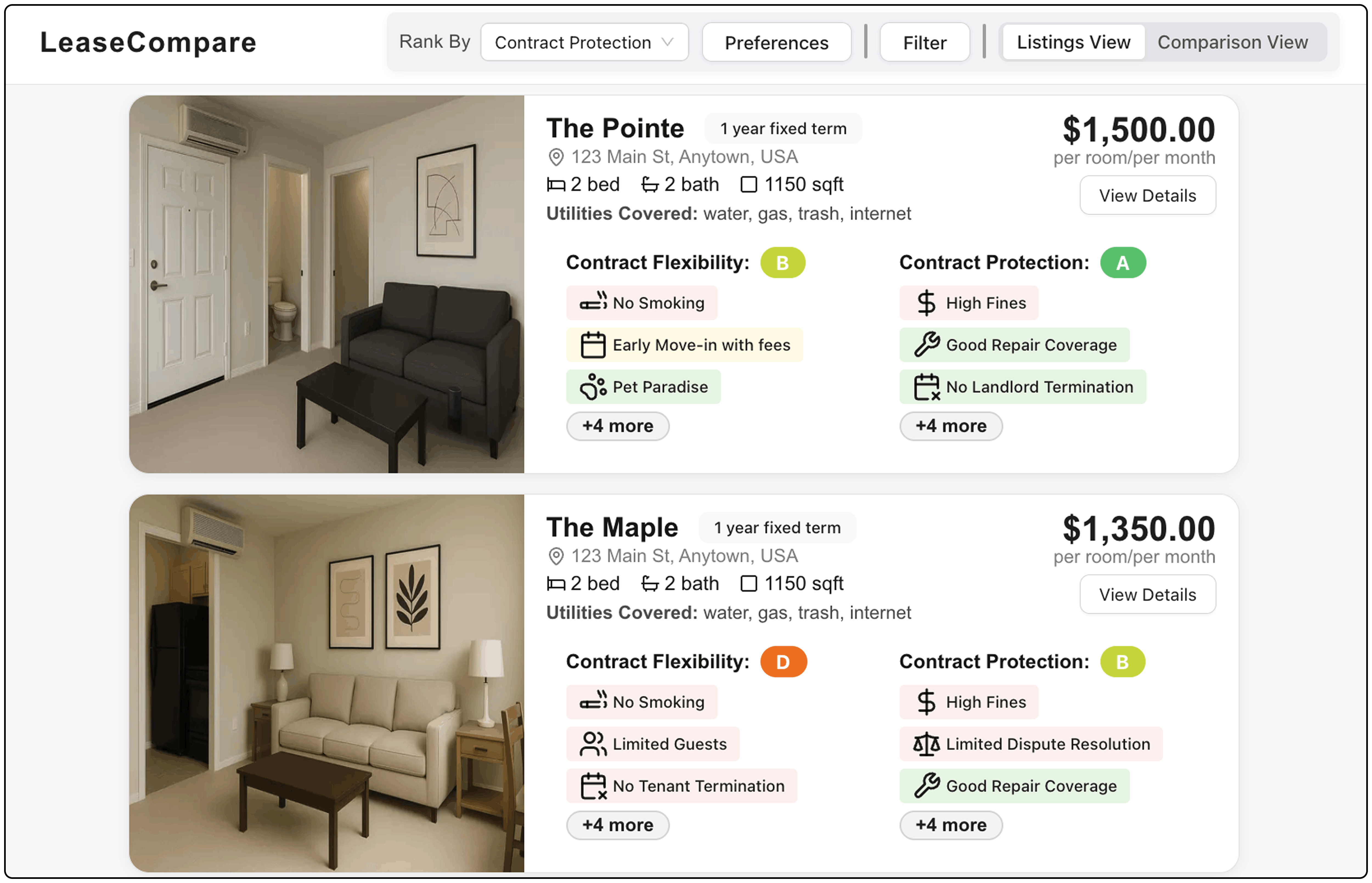}
    \caption{Listings View in LeaseCompare. Users can browse, rank, and filter apartments by contractual clauses. Each apartment features a Contract Flexibility Score and a Contract Protection Score, representing the qualities of its contract.}
    \label{fig:Leaselistingsviews}
    \Description{Listings view in LeaseCompare. At the top, a control panel allows users to rank and filter apartments not only by price but also by contractual clauses. The main screen displays apartment listings with rent, amenities, and key contractual clauses. Clauses are grouped into two categories: contract flexibility and contract protection. Each listing also shows a Contract Flexibility Score and a Contract Protection Score.}
\end{figure}

\begin{figure*}[!h]
    \centering
    \includegraphics[width=0.92\textwidth]{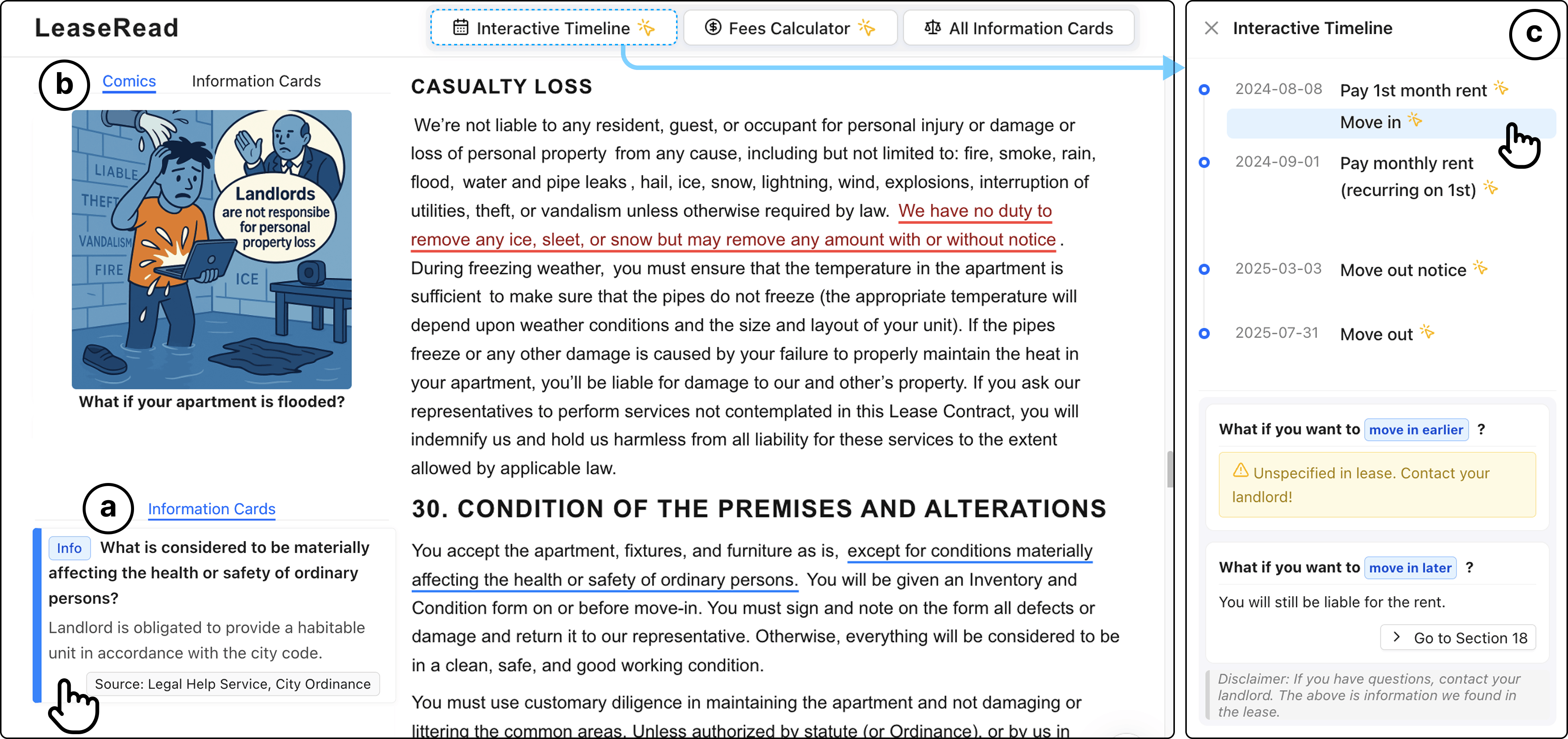}
    \caption{LeaseRead is composed of: (a) Information Cards that contextualize the contract with external legal references (e.g., city ordinance) and social comments while flagging clauses that are vague or unlawful, (b) Comic Scenarios that visualize unintended consequences of contractual clauses, and (c) Explorable Scenarios that allow users to interactively explore what-if situations. Note: Red underline signals an unlawful clause (see Appendix Figure ~\ref{fig:infocards}d for the corresponding Information Card).}
    \label{fig:LeaseReadMain}
    \Description{The LeaseRead interface with the original contract at the center and Information Cards and Comics displayed on the left of their corresponding sections. (a) Information Cards contextualize the contract with external legal references and social comments while flagging clauses that are vague or potentially unlawful. The specific information card in the figure elaborate on examples of what is considered to be unlawful housing conditions; (b) The comic scenario illustrates possible consequences if water leaked in the apartement where the landlord is not responsible for personal property loss; (c) Example Explorable Scenario that visualizes obligations in a timeline and allows users to interactively explore what-if situations such as early or late move in.}
\end{figure*}

\subsection{Before Signing: LeaseCompare}
\label{subsec:leasecompare}

Shown in Figure ~\ref{fig:Leaselistingsviews}, LeaseCompare speculates an alternative housing market and apartment search experience where contractual terms are proactively foregrounded, priced, and subject to competition (proactively supplying contractual information, DC3).

\subsubsection{Foregrounding and Summarizing Contractual Clauses}
LeaseCompare foregrounds 14 types of clauses grouped into two categories: contract flexibility (e.g., pet policies, early termination) and contract protection (e.g., rule changes, dispute resolution). Each clause is color-coded based on whether it favors the tenant, landlord, or is neutral. 
Similar to existing composite scores that summarize neighborhood features (e.g., Transit Score~\cite{transitscore}), LeaseCompare provides \textit{Contract Flexibility Score} and \textit{Contract Protection Score} to summarize the flexibility and legal protection the contract offers to the tenant. 
To explore how users might value lease clauses differently, LeaseCompare allows 
users to customize the relative importance of different clauses in contributing to the calculation of composite scores (Appendix Figure ~\ref{fig:LeaseComparecontrol}). 

\subsubsection{Comparing, Filtering, and Ranking Apartments by Contractual Clauses}
In LeaseCompare, users can browse apartments with a summary of price, amenities, and contractual clauses in the Listings View (Figure ~\ref{fig:Leaselistingsviews}), with more details in the Details Views (Appendix Figure ~\ref{fig:DetailView}).
Illustrating DC2 (i.e., alternative representation of contractual information), the comparison view aggregates and structures clauses from multiple leases into an interactive comparison table (Appendix Figure ~\ref{fig:CompareView}). 
Reminiscent of prior work that embeds democratic values into social media algorithms ~\cite{valuesocialchi}, LeaseCompare embeds contractual clauses in the apartment search and ranking algorithm. 
Users can filter out apartments containing clauses they wish to avoid (Appendix Figure ~\ref{fig:LeaseComparecontrol}) or rank apartments through Contract Flexibility or Protection Scores.


\subsection{While Signing: LeaseRead}
\label{subsec:leaseread}

While reading and negotiating a contract, LeaseRead explores opportunities for restructuring (i.e., exploring alternative representations, DC2) and contextualizing (i.e., adding external legal and social information, DC1) legal text (Figure ~\ref{fig:LeaseReadMain}).

\subsubsection{Comic and Explorable Scenarios}

LeaseRead transforms static legal text into two types of representations to support reflection and sensemaking (DC2).

\begin{itemize}
\item \textbf{Comic Scenarios:} Comic Scenarios ground and embody abstract contractual clauses in hypothetical situations (Appendix Figure ~\ref{fig:comicscenarios}). While prior work focused on using comics to summarize information (e.g., ~\cite{ketola2024comic, tabassum2018increasing}), the comic scenarios in LeaseRead are designed to provoke reflection on the unintended consequences of legal text. 

\item \textbf{Explorable Scenarios:} Explorable Scenarios aggregate and restructure information dispersed across the contract, enabling users to interactively simulate possible conditions and outcomes (Appendix Figure~\ref{fig:explorablescenarios}). 
All the Explorable Scenarios are deeply linked to the original text, meaning that as users interact with the scenarios, the original contract text that informed the outcome is dynamically highlighted.

\end{itemize}

\subsubsection{Information Cards}

To contextualize legal text to aid interpretation (DC1), LeaseRead included 32 Information Cards that provide information drawn either from external legal sources or social comments (Appendix Figure ~\ref{fig:infocards}).

\begin{itemize}

\item \textbf{Legal Information (Blue, Yellow, Red):} To educate legal knowledge beyond the contract itself, LeaseRead links contractual clauses to external legal references (i.e., housing regulations, city ordinances, university legal resources). Blue cards supply extra information, yellow cards flag clauses that may require clarification or revision, and red cards warn of legal violations.

\item \textbf{Enforcement Information (Purple):} To help users reflect on how contractual clauses are enforced in practice, LeaseRead links clauses to (hypothetical) resident comments that provide lived examples of how landlords have enforced or deviated from the contract.

\end{itemize}

\subsection{After Signing: LeaseTrack}
\label{subsec:leasetrack}

After a lease is signed, LeaseTrack probes the opportunities and concerns of proactively surfacing and reminding signees of their obligations and rights (DC3) based on \textbf{time} (Oligation Tracker), \textbf{disputes or incidents} (Incident Tracker), and \textbf{the physical environment} (Contract Aware Devices).

\subsubsection{Obligation Tracker}

\begin{figure}[!ht]
    \centering
    \includegraphics[width=0.46\textwidth]{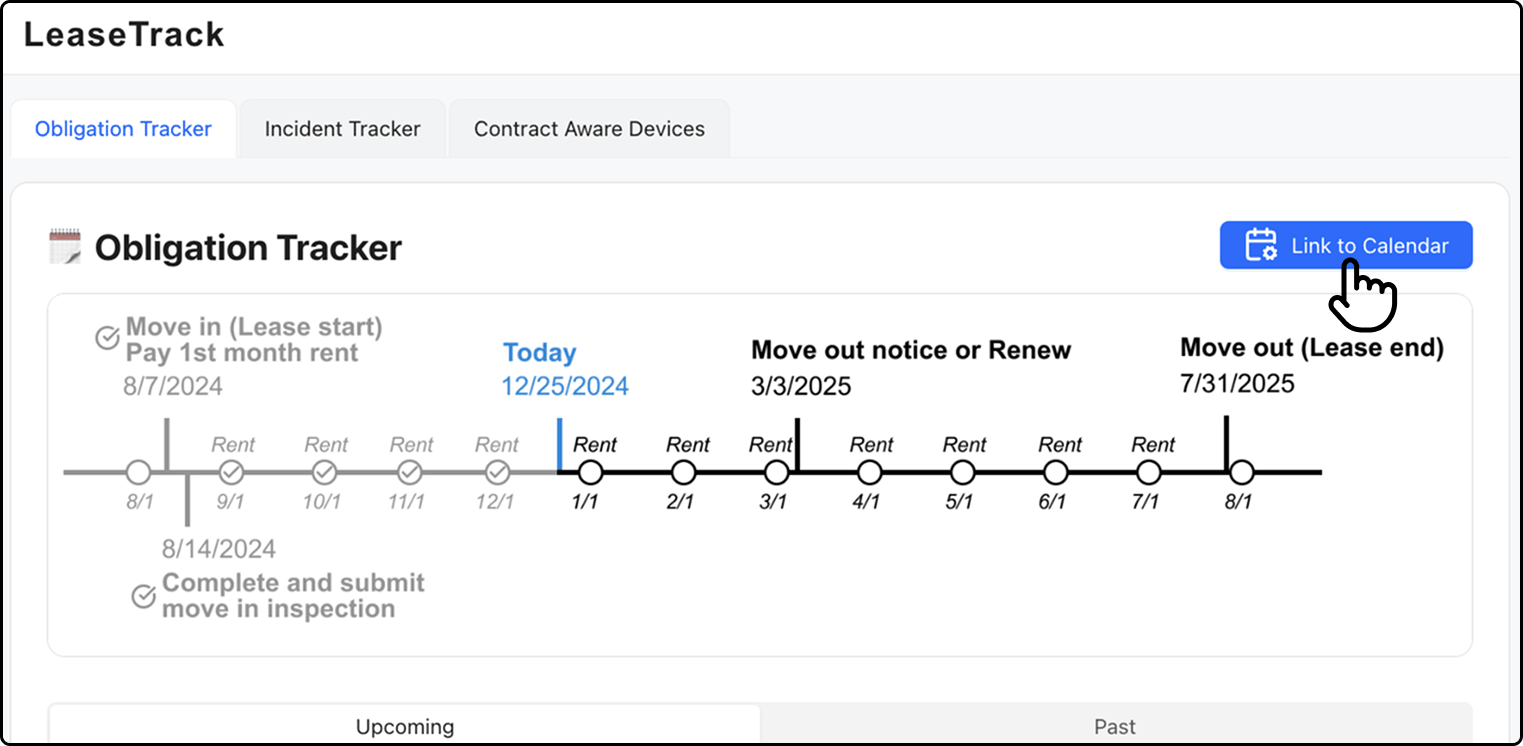}
    \caption{Obligation Tracker extracts time-sensitive contractual obligations into an interactive real-time timeline.}
    \label{fig:obligationtracker}
    \Description{Obligation Tracker in LeaseTrack. Extracts and visualizes time-sensitive obligations in a contract into a real-time timeline that updates over time. The interface allows users to view more information for each obligation or choose to link it to their calendar or email.}
\end{figure}

Shown in Figure ~\ref{fig:obligationtracker}, Obligation Tracker represents and tracks contractual obligations as a timeline (DC2), linked to a signee's email and calendar to offer proactive reminders based on time (DC3).
In the probe, clicking an obligation reveals the original clause, and the ``Link to Calendar'' button displays an example reminder email.

\subsubsection{Incident Tracker}

Incident Tracker probes the opportunities in proactively helping signees assert their rights when lease-related incidents occur (DC3). We use a scenario in which a landlord demands a late fee that may exceed legal or contractual limits. Incident Tracker flags potential rights violations, highlights relevant lease terms and local housing laws, and suggests issues or questions the tenant may wish to raise. Lastly, it offers a drafted email to aid communication with the landlord (Appendix Figure ~\ref{fig:incidenttracker}).

\subsubsection{Contract Aware Devices}
\begin{figure}[!ht]
    \centering
    \includegraphics[width=0.47\textwidth]{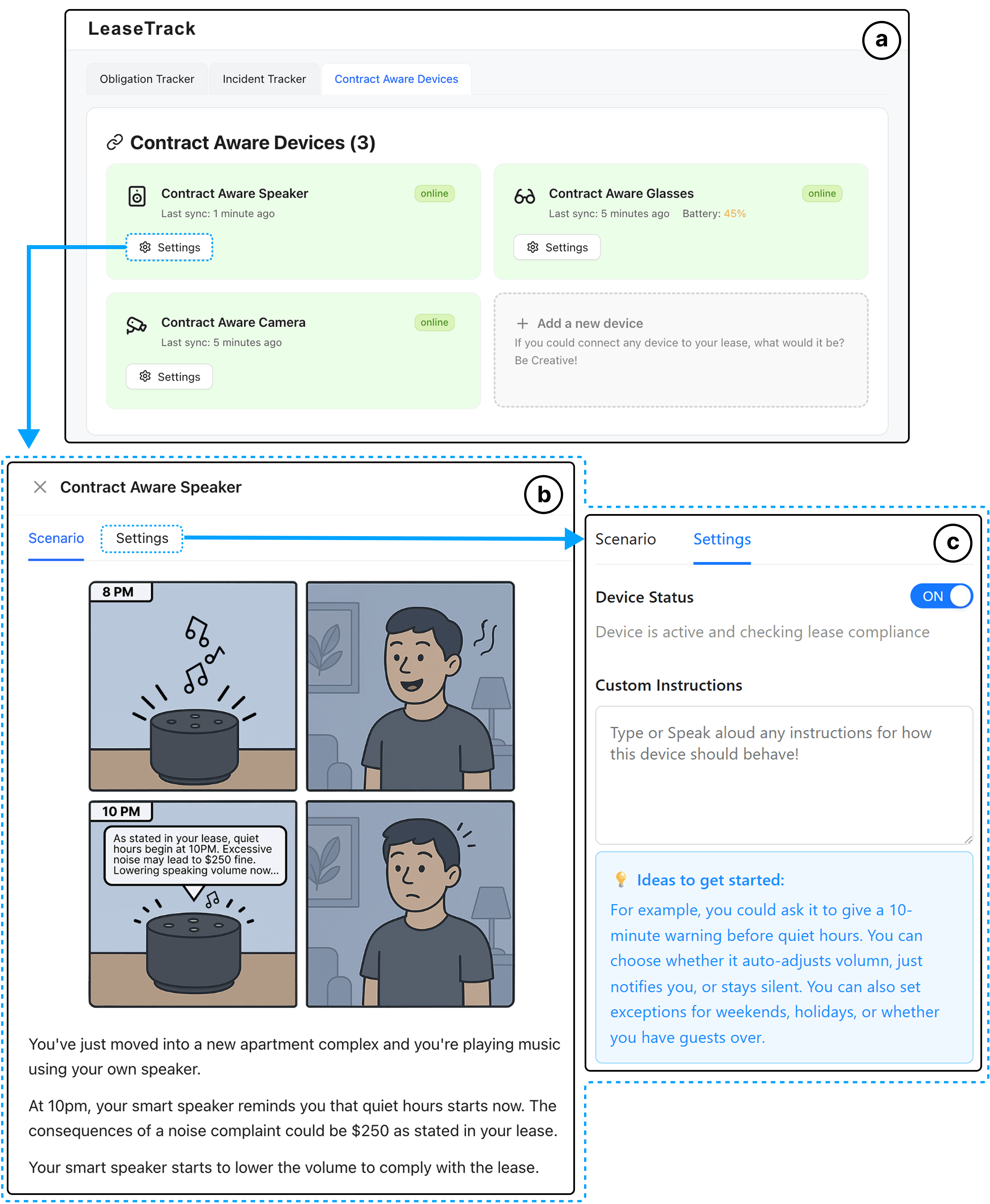}
    \caption{Contract Aware Devices: (a) three speculative smart devices that are aware of the contract and the physical environment; (b) the narrative and comic illustration of Contract Aware Speaker; and (c) the settings page that allows users to enter custom instructions on desirable device behavior.}
    \label{fig:linkeddevices}
    \Description{Contract-Aware Devices page in LeaseTrack. (a) The main page shows three speculative smart devices that are aware of both the contract and the physical environment, with a settings button to view more details for each which navigates to (b,c); (b) A narrative and comic illustration of Contract-Aware Speakers that automatically lower volume after 10 p.m. to comply with quiet-hours clauses in leases; and (c) A settings page that allows users to enter custom instructions for how the devices should behave.}
\end{figure}

Contract-Aware Devices (Figure ~\ref{fig:linkeddevices}a) explores the opportunities and concerns that arise when contracts are embodied in the environments they govern (DC3). Inspired by critical design~\cite{bardzell2011interaction, teyssier2021eyecam}, we designed three fictional devices, illustrated through comics, that provoke speculation on a future where contractual obligations have a material presence in daily life.

\begin{itemize}
\item \textbf{Contract Aware Speaker:} Detects late-night noise, reminds signees of quiet hours in the lease, and offers to automatically lower volume (Figure~\ref{fig:linkeddevices}b).

\item \textbf{Contract Aware Glasses:} Analyzes the physical environment to surface obligations and rights, such as cleanliness standards or mold growth (Appendix Figure~\ref{fig:linkedglassesandcamera}a).

\item \textbf{Contract Aware Camera:} Monitors entry attempts, alerts signees to unauthorized landlord access, and reminds signees of guest stay limits (Appendix Figure~\ref{fig:linkedglassesandcamera}b).

\end{itemize}

Signees can configure which devices are active and enter custom instructions on device behavior in the interface (Figure~\ref{fig:linkeddevices}c) to reflect on how these speculative devices should behave. 

\subsection{Study Procedure}
\label{sec:method}
We conducted a three-part study using the probes with 18 participants, guided by the following research questions:

\begin{itemize}
\item \textbf{RQ1:} What challenges did participants face in their past interactions with contracts? 

\item \textbf{RQ2:} What opportunities and concerns do participants see in the three probes illustrating Living Contracts? 

\item \textbf{RQ3:} How might the concepts illustrated in the three probes extend beyond leasing contracts?
\end{itemize}

\subsubsection{Study Procedure}

\begin{figure}[!ht]
    \centering
    \includegraphics[width=0.47\textwidth]{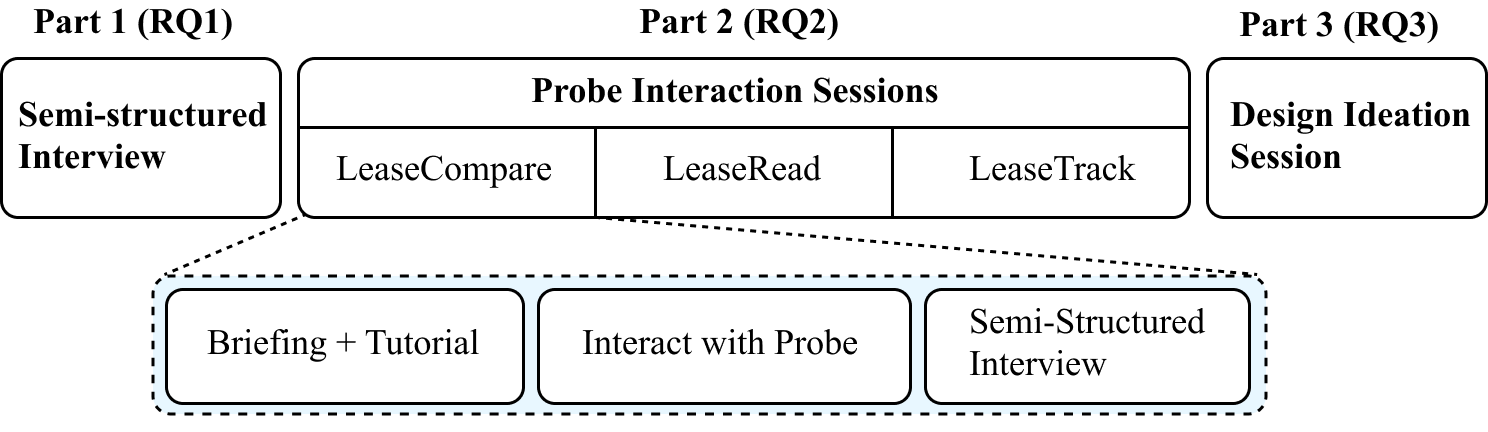}
    \caption{Procedure of the 3-part study corresponding to the 3 RQs:
    (1) initial semi-structured interview about participants' past challenges, (2) three probe interaction sessions, and (3) design ideation session. At the bottom, we show the process for each probe interaction session.}
    \label{fig:studyprocedure}
    \Description{A flowchart of the study procedure. At the top, we show an overview of the 3-part study: (1) semi-structured interview about participants' past experiences, (2) three probe interaction sessions, and (3) design ideation session. At the bottom, we show the process diagram for each probe interaction session: briefing with narrative scenarios, providing an interface tutorial, participant interaction with the probe, and a follow-up semi-structured interview.}
\end{figure}

The entire study lasted for two hours and was conducted virtually. The study was approved by our institution’s IRB, and participants were compensated with \$30. Figure ~\ref{fig:studyprocedure} illustrates the overview of the study, consisting of three parts.
To answer RQ1, we conducted an initial semi-structured interview about participants’ past challenges with contracts (Appendix ~\ref{Appendix:interviewquestionsbarriers} lists the interview questions).
Next, we introduced the study’s narrative scenario, in which participants took on the role of Alice, a student who is going through the leasing process (Appendix \ref{subsec:narratives}). 

The second part of the study consisted of three probe interaction sessions to explore RQ2. For each probe interaction session, we first introduced the narrative scenario for the specific probe (Appendix \ref{subsec:narratives}) and provided a tutorial for the probe. Participants were given 10–15 minutes to interact with the probe while thinking aloud. Afterwards, we conducted a semi-structured interview about participants' perception of the probe. 
In the study, we emphasized to the participants that the probes were fictional and that our aim was to elicit imagination and reflection on future contract interfaces rather than to evaluate feasibility or usability ~\cite{chow2025beyond}. We also encouraged participants to critique, modify, or expand upon the ideas illustrated in the probes. The same procedure was repeated for each of the three probes.

Lastly, to explore RQ3, participants engaged in a design ideation session. Participants were asked to choose and augment one or more contract types other than leases that they had interacted with. Participants were encouraged to draw inspiration from the three probes or propose new ideas. 

\subsubsection{Participants}

In total, 18 participants took part in the study. We recruited participants through Prolific. All were over 18 years old, located in the United States, and had signed both a leasing contract and at least one other type of contract. 
We coded participant interviews after every 5 interviews, and continued recruiting until reaching data saturation (i.e., no new codes for the 5 previous interviews).
All participants were fluent in English, and one participant self-identified as a non-native speaker (P5). 
One participant was a landlord with experience reviewing and drafting leases (P4), four described regularly interacting with contracts or legal documents as part of their work (P1, P3, P10, P14), and one had completed graduate-level coursework related to contract law (P12).
Table ~\ref{tab:demographics} summarizes participant demographics.

\subsubsection{Analysis}
We analyzed interview transcripts and screen recordings of participants’ interactions with the probes using reflexive thematic analysis, following Braun and Clarke’s six-phase framework~\cite{braun2006using}. 
Two researchers, the first and last author, began by familiarizing themselves with the data by reading through the interview transcripts independently. They each read three interviews, grouping quotes under descriptive initial codes related to the research questions. They then met to discuss these initial codes in two one-hour meetings to resolve differences and merge into a single agreed-upon set. 
They were predominantly HCI researchers with experience running mixed-methods studies focused on design, document augmentation, and human-centered AI. During these meetings, the two coders also shared findings with other members of the team, including a legal scholar who specializes in contract law. 
Using these codes, the first author then coded the remaining data, iteratively refining and adding codes as new insights emerged, and grouping related codes into themes. When adding new codes beyond the initial set, the researcher shared new codes with relevant quotes in meetings with the rest of the research team. 
The research team met weekly for a month to review and refine the final codes and themes, ensuring that interpretations were informed by multiple disciplinary perspectives (i.e., HCI and law). In line with a reflexive thematic analysis approach, we prioritized multivocality over consistency and therefore do not report inter-rater reliability ~\cite{mcdonald2019reliability, forbes2022thematic}. 
Below, we present findings from our analysis.



\section{Findings}

We organize our findings by our three research questions, aligning with participants' responses from the three parts of the study.
In \ref{subsec:RQ1findings}, we report the challenges participants encountered with contracts elicited from the initial semi-structured interview (RQ1). In \ref{subsec:findingsrq2}, we report participants' perceptions and reflections around the three design probes during the probe interaction sessions (RQ2). Lastly, in \ref{subsec:participantideas}, we report participants' ideas for augmenting contracts beyond leases during the design ideation session (RQ3).

\subsection{RQ1: What challenges did participants face in their past interactions with contracts?}
\label{subsec:RQ1findings}

First, we report findings from the initial semi-structured interview (Study Part 1) on participants' challenges with contracts such as leases, insurance policies, rental contracts, loans, mortgages, medical agreements, and employment contracts (Summarized in Table~\ref{table:readingbarriers}).

\begin{table*}[h!]
\centering
\renewcommand{\arraystretch}{1.5} 
\setlength{\tabcolsep}{5pt} 
\begin{tabular}{p{4cm} p{10cm} p{1cm}}
\hline

\textbf{Barrier} & 
\textbf{Example Quote} &
\textbf{Count} \\
\hline

Contracts not available early &  
\textit{``you usually don't get to see the contract before you're right about to sign.''} (P6) &
10 \\
\hline

Difficulty comparing contracts &  
\textit{``I mean how many times can you divide your screen to be able to compare them. Nobody's gonna do all that, it's too much ...''} (P3) &
13 \\
\hline

Extended and dense text &  
\textit{``There's a lot of information in a contract, so it can get a little bit overwhelming. And then they like to hide in the small print.''} (P7) &
13 \\
\hline

Legalese and jargon &  
\textit{``the information was so confusing ... and they use such jargon. That you don't know half of what you're reading.''} (P7) &
15 \\
\hline

Vague and strategic language &  
\textit{``Sometimes it’s vague ... Most of your tenant protections are found in law rather than in leases.''} (P4) &
15 \\
\hline

Difficulty interpreting or being aware of the law &  
\textit{``Nobody understands the law. I mean, they don't make sense.''} (P3) &
13 \\

\hline

Language barriers as non-native speakers &  
\textit{``It's full of all this unusual type wording, especially for me, because English is not my first language.''} (P5) &
1 \\
\hline

Lack of experience when signing at a young age &  
\textit{``I did my student loan at 19, and it was a lot of things I didn't understand... my mother didn't have any knowledge about it either.''} (P10) &
8 \\
\hline

Physical pain when signing a surgical contract &  
\textit{``Physically, when you're in that kind of headspace you can't think straight...In my case, that was an extreme amount of pain ...''} (P7) &
1 \\
\hline

Feeling pressured and rushed to sign &  
\textit{``Oftentimes you're rushed to make a decision... It's just a quick glance over before deciding whether or not to accept it.''} (P15) &
7 \\
\hline

\hline
\end{tabular}
\caption{Participants' self-reported barriers in the initial semi-structured interview (RQ1).}
\label{table:readingbarriers}
\Description{Participants’ self-reported barriers to interacting with contracts. Barriers include: contracts not available early (10), difficulty comparing contracts (13), extended and dense text (13), legalese and jargon (15), vague and strategic language (15), difficulty interpreting or being aware of the law (13), language barriers as non-native speakers (1), lack of experience when signing at a young age (8), physical pain when signing a surgical contract (1), and feeling pressured and rushed to sign (7).}
\end{table*}

\subsubsection{\textbf{Participants engaged with contracts mainly at the point of signing}}

Sixteen participants noted that contracts were not always considered early in their decision-making process. For example, they primarily relied on amenities and rent to make housing decisions, engaging with leases only at the point of signing. In many cases, the contract was not available until late in the process (10/18). Even if the contracts were available earlier, participants described the difficulty in surfacing important clauses and comparing multiple contracts due to their length and complexity (13/18).
Consequently, participants explained that there's often sunk cost before they actually review the contracts (e.g., searching and applying for apartments), at which point they are reluctant to reject a transaction because of its contract.  
\begin{quote}
    \textit{``You may not know so many things until you are actually signing off on the contract. Then, it's like you can't go back, or you don't want to go back because you've been so much through that process of finding something.''} (P10)
\end{quote}

\subsubsection{\textbf{Participants struggled to parse extended and dense legalese}}

Most participants reported that they try to read (12/18) or at least skim the contract (4/18) for higher-stakes settings (e.g., housing, insurance, medical, etc).
Yet, important details were often buried within a wall of text that discouraged engagement (13/18).

\begin{quote}
    \textit{``My lease was probably like 40 pages. You wouldn't have known where to find the important details. Then, people get frustrated, or they just feel like they don't have enough time to do it or deal with that.''} (P9)
\end{quote} 

Participants also emphasized how the legalese and jargon made it difficult to follow the logic, often requiring repeated reading to piece together the intended meaning (15/18).
\begin{quote}
    \textit{``When it's dealing with legalese-type language, it can get pretty confusing. It takes a while for you to track and backtrack… read the entire thing, and then pull it apart and be like, I think this is what they mean.''} (P11)
\end{quote}

\subsubsection{\textbf{Participants lacked awareness of their legal rights}}

Fifteen participants described encountering strategic language that was intentionally vague or misleading, making it difficult to understand their legal rights.
For example, P4, a landlord, acknowledged that tenant protections are often omitted from leases and that vagueness may be deliberate to maximize landlord protection.

\begin{quote}
    \textit{``Sometimes it's vague, and it’s vague on purpose. Most leases are like that. The lease is really set up more to protect the landlord. Most of your tenant protections are found in law rather than in leases.''} (P4)
\end{quote}

However, interpreting the nuances of law can be difficult, especially since laws vary across jurisdictions, change over time, and contain inherent ambiguity, leaving participants unaware of and vulnerable to unlawful requests (13/18).
This challenge can be compounded by the common difference in power between the service provider and consumer, which made participants reluctant to challenge the bigger party. 

\begin{quote}
    \textit{``People think they (service provider) can do a lot of stuff. But, some of the time, they can't really do it. Just the intimidation factor of it. Especially if it's like a big corporation, or entity that's bigger than myself, I'm assuming that what they're doing is legal.''} (P11)
\end{quote}

\subsubsection{\textbf{Circumstances defined extent of participant's engagement with contracts}}
\label{findings:situationalbarriers}

Participants described being in situations where they were unprepared to meaningfully engage with the contract, such as signing leases or student loans at a young age (8/18), facing language barriers as a non-native speaker (P5), or experiencing physical pain when signing a surgical contract (P7).
Seven participants also reported challenges associated with interpersonal dynamics, feeling pressured by the other party to sign quickly to move the process forward.
\begin{quote}
    \textit{``There's a lot of time pressure, and they [landlords] are waiting on you to sign it [the lease] so they can give you the keys. so that they can get on with their day.''} (P6)
\end{quote}

Relatedly, P1, a mortgage banker, and P4, a landlord, observed that first-time buyers or renters were often so excited about a new home that they did not pay attention to contracts, especially when facing persuasive sales tactics.
\begin{quote}
    \textit{``They're excited about this beautiful new home, their future, and their family. A lot of times, they overextend themselves because of unscrupulous real estate agents. They keep up-selling them. They're letting their emotions guide them rather than rational thoughts.''} (P1)
\end{quote}

\subsection{RQ2: What opportunities and concerns do participants see in the three probes illustrating Living Contracts?}
\label{subsec:findingsrq2}

During the probe interaction sessions (Study Part 2), participants interacted with three design probes. Below, we report the themes of participants’ reflections and feedback on future contract interfaces.

\subsubsection{\textbf{Participants perceived contracts as interfaces that facilitate relationships}}
\label{sebsec:ContractAsRelationship}

When comparing listings with LeaseCompare, all participants were willing to pay more for a lease with more tenant-friendly clauses. 
This was reflected in the apartments they chose for their character, Alice: 16 participants selected the apartment with the highest rent and most tenant-favorable clauses (\$1,500), 1 selected the mid-priced option (\$1,350), and 1 described preferring to negotiate before deciding.
15 participants described evaluating contractual clauses in terms of potential financial risks. For example, P5 valued a clause that limited her liability to only her share of the rent. P5 explained that if her roommate stopped paying, she would otherwise owe much more, and described the clause as \textit{``protecting mental peace''}. Participants also perceived the contract as a \textit{``psychological interface''} (P4) that reflects the potential relationship between the parties. Reflecting this, 17 participants described prioritizing listings with clauses that signaled a sense of autonomy (6) and balance of power (16), even if the likelihood of invoking them is low.

\begin{quote}
\textit{``Even though in most cases you're not gonna need to exercise this [dispute resolution through court rather than mandatory arbitration], it gives more power to the lessee. I think it would also make the landlord hesitant to do anything unscrupulous or unfair.''} (P15)
\end{quote}

\subsubsection{\textbf{Knowledge not in the contract could help consumers advocate for themselves (DC1: Contextualization)}}

When interacting with LeaseRead, all 18 participants emphasized that in-situ legal references in the Information Cards (e.g., explanations of city ordinances) taught them about legal rights they were previously unaware of.
Moreover, the Information Cards helped participants surface ambiguous and potentially unlawful clauses, as well as past incidents they wouldn't have identified on their own (18/18).
Equipped with this information, participants felt more empowered to request clarifications and modifications, especially under time or social pressure.

\begin{quote}
\textit{``It's the awareness and the legal knowledge [that makes interpretation hard]. And then there's that time pressure ... They want you to just blindly accept. These information cards give more power to the signer to be able to say: hey, wait a minute. This needs clarification. I want this modified in writing for me.''} (P6)
\end{quote}

\subsubsection{\textbf{Alternative representation of legal text could aid challenging information tasks for consumers (DC2: Representation)}}

Throughout the three probes, participants highlighted the opportunities of different contract representations in supporting decision making (Comparison View), sensemaking (Explorable Scenarios), reflection (Comic Scenarios), and obligation management (Obligation Tracker).
First, participants valued LeaseCompare for making contractual information more transparent by extracting key clauses that would otherwise be buried in text (14/18). 
Paired with the comparison view, LeaseCompare enabled side-by-side evaluation of multiple contracts, an information task participants described as previously impossible (14/18).

In LeaseRead, participants described Explorable Scenarios as having the potential for \textit{``eliminating miscommunication and misunderstanding''} (P10).
For example, 17 participants explained how the interactive timeline and fees calculator visualized scattered information that they otherwise might miss. Moreover, the Explorable Scenarios simplified dense legal text by allowing participants to interactively explore contractual consequences (16/18).

\begin{quote}
\textit{``Lease language really becomes much easier by walking me through the interactive way. It immediately lets me play with different situations and think about all kinds of scenarios. It pulls everything together.''} (P5)
\end{quote}

The Comic Scenarios further presented opportunities for attracting attention (13/18) and prompting reflection on the unintended consequences and underlying intentions of contractual clauses (9/18).
At the same time, two participants, including the landlord, cautioned that the provocativeness of the comics may risk producing inaccurate portrayals of the service provider.

Lastly, 10 participants valued LeaseTrack's Obligation Tracker as a centralized, real-time timeline for monitoring obligations rather than relying on memory or repeatedly consulting the original lease. P7 specifically noted its potential for managing multiple contracts simultaneously.

\subsubsection{\textbf{Proactively supplying contractual information could be empowering, but held risks (DC3: Proactivity)}} 

Participants noted how proactively supplying contractual information could raise their legal awareness and help them make more informed decisions before (e.g., selecting an apartment in LeaseCompare: 17/18) and after signing (e.g., managing obligations or advocating for rights in LeaseTrack: 18/18).
Consequently, participants described how they would feel more \textit{``empowered to speak from a place of knowledge in conversations with the landlord''} (P2) when clarifying or negotiating contractual clauses before signing (9/18) and advocating for themselves upon unlawful requests after signing (16/18), especially in situations when they might not be able to critically engage with contracts.
A few participants envisioned the broader market effects. For example, foregrounding contractual clauses before signing (e.g, LeaseCompare) may incentivize landlords to compete for better clauses (5/18) while proactively reminding users of their rights after signing (e.g, Incident Tracker) may discourage landlords from making unlawful requests (2/18).

Participants' responses were more nuanced for Contract-Aware Devices that embed lease awareness into the home.
16 participants saw opportunities for Contract-Aware Devices on avoiding accidental contract violations, supporting people with cognitive disabilities (e.g., Alzheimer’s), reminding guests or children, or preventing unlawful landlord actions (e.g., unlawful entry).
At the same time, 13 participants worried about device data access by the service providers or governments.
Eight participants preferred devices that only issue reminders, noting that auto-enforcement (e.g., a speaker auto-lowering volume during quiet hours) could feel \textit{``creepy and more like an enforcer than an assistant''} (P2). 
P7 further noted a preference for text reminders over anthropomorphic voice alerts to reduce the sense of being monitored.
Finally, some participants envisioned that constant reminders may diminish the sense of home (10/18), and contractual clauses may not be meant to be uniformly enforced in practice (9/18).

\subsubsection{\textbf{From the other side: values and concerns of Living Contracts with service providers}}

One of our participants, P4, was a landlord. While in some cases P4 aligned with how other participants responded to the probes, P4 also provided interesting counterpoints that highlight some of the tensions that our probes touched upon. 
First, the probes illustrated compelling alternatives to the landlord's current practice. 
P4 noted how many of his tenants do not pay attention to leases. Consequently, P4 highlighted the value of features that could inform tenants to be more conscientious (e.g., LeaseCompare, Explorable Scenarios in LeaseRead). 
Further, offering tenants the ability to filter apartment searches by lease terms might also select for tenants who were more attentive, reflecting a similar attitude as other participants in viewing contracts as a potential relationship (\S\ref{sebsec:ContractAsRelationship}).

\begin{quote}
\textit{``The people who would pay attention to [LeaseCompare] are the kind of tenants I'd want. If they're going to pay that much attention to the details in the lease
and they're going to pick at me about how long guests are welcome, or
what the sublet fees are... they're probably going to take better care of my place.''} (P4)
\end{quote}

While P4 valued the convenience of systems that could automatically remind tenants of their obligations after signing (e.g., Obligation Tracker, Contract-Aware Devices), he cautioned that machines may only be able to make rigid decisions based on the contract. He prefers managing reminders himself to be more humane, making accommodations based on tenants' situations. {For example, P4 described that he might not charge a late fee to tenants who recently suffered from a family loss and had to pay for funeral.

P4 was also ambivalent about features that focused on tenant rights. On one side, he expressed concern that features may help tenants uncover flaws in his lease or requests (e.g., Information Cards, Incident Tracker). At the same time, P4 emphasized that these features could help him validate his lease and requests. For example, P4 recalled his own experience being confused by the vagueness of local law when drafting leases and requesting fines (i.e., landlords can only charge fines that are \textit{``fair and reasonable''}).

\subsection{RQ3: How might the concepts illustrated in the three probes extend beyond leases?}
\label{subsec:participantideas}

Table~\ref{table:participantdesignideas} provides an overview of participants' ideas in the design ideation session (Study Part 3), illustrating how features of the probes and the design concepts of Living Contracts may inspire interfaces for contracts beyond leases.
Below, we present two case studies of participants' ideas.

\subsubsection{\textbf{Case Study 1: Multi-Contract Explorable Scenarios}}

Reflecting on past experience with surgical contracts, P7 explained that key costs were often buried in dense text, causing her to estimate the wrong surgical cost. She noted a similar challenge with her medical insurance contract, leading her to pay for procedures she later realized were covered.
Inspired by LeaseRead, P7 proposed a new type of Explorable Scenario that integrates both surgical and insurance contracts to estimate costs and coverage.
\begin{quote}
\textit{``For interactive scenarios, in addition to calculating the fee based on the number of nights [in the hospital], it will be useful to have a feature that allows you to see whether you're covered or not and what your rights are. There could be a variation based on how much medication you need.''} (P7)
\end{quote}

P7’s idea demonstrated how the concept of representing legal text into Explorable Scenarios (DC2: Representation) may go beyond leases to span multiple contracts. 





\subsubsection{\textbf{Case Study 2: Contract Aware Devices for Emergency}}

P2 proposed contract-aware devices for post-accident decision-making, when emotions often override rational thought. She envisioned a car camera that could detect an incident and remind drivers of the steps needed to secure coverage.

\begin{quote}
\textit{``If someone is getting into a car accident, that's such a traumatic and jarring experience. What you do immediately after an accident has such heavy bearing on whether you are going to get compensation afterwards, who's liable, and all these different things. It can be hard to think about all those things in a moment. If the technology is acting as that logic person in the room, it can be extremely helpful to make sure that you're covering all your bases.''} (P2)
\end{quote}

P2’s proposal demonstrated how contract-aware devices that proactively supply contractual information may provide value in high-stress, emergency settings (DC3: Proactivity).



\section{Discussion}

In this work, we explored the challenges of interacting with contracts and the opportunities for new contract interfaces to tackle these challenges. We present three design probes based on supporting contract engagement before, during, and after signing. Our findings highlight both consumers' and providers' excitement about such designs. Below, we discuss the implications of our findings around how contracts shape relationships between parties (\S\ref{subsec:disc:relationalinterface}), the social and contextual challenges contracts present (\S\ref{subsec:disc:socialContext}), and opportunities for moving beyond static documents towards Living Contracts (\S\ref{discussion:oppertunities}).

\subsection{Contracts as Relationships}
\label{subsec:disc:relationalinterface}

While contracts are often framed as legal instruments for dispute resolution~\cite{kar2019pseudo, haapio2021contracts, nousiainen2022legal}, participants in our study also perceived them as reflecting an emerging relationship with the other party, aligning with the theory of relational contracting~\cite{macneil1985relational}.
For example, when interacting with LeaseCompare, all participants indicated a willingness to pay more for tenant-friendly clauses. Beyond evaluating financial risks, as highlighted in prior work~\cite{korobkin2003bounded, dasgupta2007lease}, participants interpreted clauses as signals of the landlord–tenant relationship. They prioritized provisions that communicated autonomy and balance of power, even when the likelihood of invoking the clauses was low (e.g., dispute resolution through court rather than mandatory arbitration). Additionally, the landlord in our study saw LeaseCompare as potentially selecting for more conscientious tenants, reflecting a similar recognition that leases represent the start of a relationship for both signing parties.  
These findings suggest that interface designers must carefully consider how new features might affect the type of relationship a contract is signaling. 
While prior work in contract design often focused on trying to evaluate and improve performance measures (e.g., comprehension)~\cite{passera2015beyond, tabassum2018increasing}, future work could investigate how different design choices influence consumers' perception of the contractual relationship, such as the sense of security~\cite{akalin2019evaluating}, trust~\cite{shneiderman2000designing}, or power~\cite{schneider2018empowerment}.

\subsection{Challenges of Human-Contract Interaction}
\label{subsec:disc:socialContext}

Prior work on high-stakes contracts has focused on the challenges associated with the document itself, such as analyzing unenforceable clauses~\cite{furth2017unexpected, hoffman2022leases, prescott2024subjective, hicks1972contractual}. Our findings suggest that the challenges people face when interacting with contracts often extend well beyond the document.
Participants encountered situational barriers, such as signing contracts at a young age, as non-native speakers, or, in one case, while in physical pain. 
Interpersonal dynamics further limited engagement. Some participants felt pressure to sign quickly and were reluctant to challenge the ``bigger'' party, especially given their limited legal knowledge. At the market level, participants often reviewed contracts late in the transaction process, after investing substantial effort and costs. At the technological level, participants described lacking tools and resources to meaningfully compare multiple contracts and interpret the nuances of law that could vary across jurisdictions. Taken together, future intelligent interfaces should recognize and take into account the situational challenges people face when interacting with contracts. 
As shown in our findings (\S\ref{subsec:findingsrq2}), Living Contracts present opportunities for empowering users to navigate these situational barriers.

\subsection{Towards Living Contracts: Opportunities and Challenges}
\label{discussion:oppertunities}

\subsubsection{Opportunities for the HCI community}

Our work offers a preliminary exploration of Living Contracts and presents several opportunities for future HCI research.
First, we invite researchers to expand the design space of contract interfaces for different types of agreements and stakeholders, drawing from the concept of Living Contracts (\S\ref{sec:designconcept}), the three design probes (\S\ref{sec:designprobes}), and participants' ideas in the design session (\S\ref{subsec:participantideas}). 
For example, governmental agencies could employ interfaces like LeaseCompare or LeaseRead to audit contracts at scale for legal compliance, rather than intervening only after lawsuits. Corporations could adopt the Obligation Tracker to manage hundreds of business contracts simultaneously. 
We plan to contribute to this design space by open-sourcing our probes as reusable templates\footnote{https://github.com/hzhfred/LivingContracts}.
These templates can either be directly reused (e.g., Comic Scenarios) or be repopulated with custom contract-specific details (e.g., LeaseCompare, Explorable Scenarios, or Obligation Tracker). 
Over time, we envision an expanded repository of reusable interface templates for different contracts and audiences, similar to existing repositories for explorable explanations in scientific publications \citep{explorexp}. 
These templates could also support contract drafters and designers.
Building on prior work in authoring tools for interactive academic papers~\cite{heer2023living, tao2025freeform}, future work could develop novel tools for authoring Living Contracts, such as helping drafters create Comic or Explorable Scenarios for their contracts. 
Lastly, we encourage researchers to empirically investigate the effects of features inspired by Living Contracts to arrive at more granular design recommendations. For instance, prior work found that abstract comics were more persuasive for charity donations~\cite{xiao2019should}. It's possible that different presentations of the Comic Scenarios in LeaseRead could elicit different reactions and effects, suggesting an interesting avenue for further study.

\subsubsection{Societal and market considerations}
In the study, participants highlighted how the concept of Living Contracts has the potential to open contracts to market forces. While exciting, deploying systems inspired by Living Contracts may also encounter barriers associated with market competition. For instance, landlords may be reluctant to disclose their leases early on a platform like LeaseCompare if they intentionally include many one-sided clauses. It might also be difficult to encourage tenants to share their renting experience (e.g., for social comments on a contract) due to privacy concerns. Furthermore, in markets with little market competition (e.g., significantly more tenants than housing available), consumers may lack meaningful choice, rendering a tool like LeaseCompare more of an informational resource than a decision-making aid. 
On the other hand, encouraging market competition on contractual clauses in LeaseCompare may unintentionally lead to housing disparity where low-resource tenants are left with predatory contracts.
We believe that any meaningful implementation of Living Contract concepts would require buy-in from multiple stakeholders (e.g., tenants and regulators) and policy-level innovation.
Past work on HCI-Policy collaboration \cite{yang2024future} and balancing multi-stakeholder expectations in public system deployments \citep{10.1145/3432908} offers a compelling start. 
For example, future work could explore policy innovations on mandating landlord disclosure of leasing contracts on the internet to be able to obtain a leasing license.

\subsubsection{Opportunities for the AI community}
While the probes we presented were not powered by LMs, they lay a path forward for the capabilities needed to implement Living Contracts and the challenges of doing so. 
First, Living Contracts would need new model capabilities to both retrieve relevant legal information and reason about the contract itself (e.g., Information Cards: Figure~\ref{fig:infocards}). 
While prior work has focused on retrieval systems for statutes and case law~\cite{yue2024lawllm, zheng2025reasoning, ryu2023retrieval} or models that interpret contracts against specific regulations~\cite{leivaditi2020benchmark, hendrycks2021cuad}, our findings highlighted the ambiguity and jurisdictional variation of law.
For future systems to navigate these nuances, a jurisdiction-specific repository of laws paired with expert interpretations is needed~\cite{mitchell2018interpretation}. Ideally, the repository is regularly updated to keep track of the latest legal regulations.

Second, our findings highlight the opportunities for alternative contract representations.
Future work can develop techniques to automatically transform legal text into alternative representations, a task with which current models struggle~\cite{barbara2024automatic}. 
Given that legal contracts are legally binding and, as our study found, represent a relationship, researchers must be cautious of the risk of misrepresentation. For example, P4, the landlord, cautioned that the Comic Scenarios in LeaseRead may risk creating inaccurate portrayals of the provider. This might require multi-stakeholder (e.g., both tenant and landlord) input on model output. While past work has explored human-in-the-loop systems for aligning model output to an individual user \citep{sorensen2024roadmappluralisticalignment, shaikh2025aligninglanguagemodelsdemonstrated}, there are additional opportunities to integrate feedback from multiple, potentially conflicting humans, into a single response.

\subsection{Limitation and Future Work}
The nature of this work is exploratory.
Similar to prior speculative design studies~\cite{chow2025beyond, noortman2019hawkeye} and Gaver's intention in using probes~\cite{gaver1999design, boehner2007hci}, we do not aim to test the usability of our probes or elicit conclusive requirements. 
Instead, this work intends to inspire and open the design space of contract interfaces for future work to build on. 
For example, future work could implement the Explorable Scenarios in LeaseRead and conduct controlled quantitative evaluations of usability, cognitive load, or comprehension.
Additionally, while our study included participants with diverse backgrounds---from non-native speakers and lay consumers to legal professionals and service providers---future work could focus on other stakeholders or user groups, such as city officials, corporations, or marginalized communities. 
For example, prior work has shown how marginalized communities may distrust the legal system due to historical inequalities ~\cite{butler1995racially, gibson2018black}, and neurodiverse readers may encounter additional barriers when making sense of complex documents ~\cite{maccullagh2017university, stern2013role}.
Lastly, our study focused on U.S. contracts and participants. Because institutional and legal traditions may vary across countries~\cite{menski2006comparative}, future work should explore how people in other countries engage with contracts and how the concept of Living Contracts may be adapted internationally.

\section{Conclusion}

In this paper, we propose Living Contracts, the idea that contract interfaces can go beyond static document readers to become interactive interfaces that adapt to different information tasks before, during, and after signing. 
Using residential leases as a case study, we designed three interactive probes that represent possible Living Contract concepts throughout the renting process.
Our three-part qualitative study (N=18) revealed situational barriers that extend beyond the contract itself, such as late access to contractual information and pressures to sign. Participants' feedback on the probes highlighted the opportunities of Living Contracts in providing in-situ legal education, transforming contractual information into task-specific representations, and proactively surfacing contractual information to aid decision-making and rights advocacy. Participants further brainstormed how the vision of Living Contracts may extend beyond leases. Taken together, our findings open up a new design space for human-contract interaction.


\bibliographystyle{ACM-Reference-Format}




\appendix


\section{Design Process and Implementation Details}
\label{Appendix:additionaldetails}

All three probes are implemented as interactive web applications using Next.js. Below, we describe the design process of the probes (\ref{subsec:probeconsiderations}), additional details for each probe (\ref{Appendix:LeaseCompare}, \ref{Appendix:LeaseRead}, \ref{Appendix:LeaseTrack}), and the narrative scenarios we designed to ground participants' interaction with the probes (\ref{subsec:narratives}).

\subsection{Design Process and Considerations}
\label{subsec:probeconsiderations}
In line with prior speculative design literature using probes, the goal of the probes is not to evaluate interface usability, but to provoke imagination and reflection on future interfaces for legal contracts ~\cite{chow2025beyond, noortman2019hawkeye}. Below, we describe the design process our research team undertook.

\subsubsection{Design Process.}
\label{subsubsec:designprocess}

To explore the idea of going beyond document-centric interaction with contractual information, we take initial inspiration from prior theories in HCI and Law (e.g., papers referenced in ~\ref{sec:designconcept}). 
Together with multiple discussions with HCI and Legal Scholars within the research team, we synthesized four design concepts: Contextualization, Malleable Representation, Proactivity, and Embodiment. Here, Embodiment refers to the ability for contract interfaces to be aware of the physical environment (e.g., Contract-Aware Devices). We decided to combine Proactivity and Embodiment as one design concept because these they are inter-related (i.e., to be able to offer proactive intervention, the system has to be situation-aware). 

While these prior theories and concepts inspired our ideation, we weren't constrained by them when ideating the probes. Using leasing contracts as a case study, we then engaged in multiple open-ended ideation sessions spanning the course of 3 months in which the research team generated, iterated on, and mocked up over 30 design ideas in Figma, spanning early apartment search, contract reading, and obligation management. Afterwards, we thematically categorized our design ideas, and the three design concepts ended up summarizing the idea space well.

Using the design mockups, we conducted three pilot studies with prior tenants (2 hours each) to explore the probes’ potential to prompt reflection and speculation. The pilot studies followed a structure similar to prior work adopting mockup-assisted interviews ~\cite{kambhamettu2024explainable}, beginning with semi-structured interviews about past challenges, followed by participants reflecting on the design mockups. Below, we outline the changes that we made based on feedback from the pilot studies.

\subsubsection{Consideration 1: Making probes interactive and grounded in scenarios}
In our pilot studies, participants reported having difficulty experiencing and reflecting on the static mockups in depth. In response, drawing on prior work with interactive design probes~\cite{salovaara2025triangulating, noortman2019hawkeye}, we implemented each probe as an interactive web interface accompanied by narrative scenarios (Appendix \ref{subsec:narratives}) to situate the probes in a lived context. 
For example, LeaseCompare allows participants to interactively compare apartment listings by contract (\S\ref{subsec:leasecompare}), and LeaseRead allows participants to interact with explorable scenarios (\S\ref{subsec:leaseread}).

\subsubsection{Consideration 2: Inviting participant input and reflection}
We noticed that while the feature mockups could illustrate specific design ideas, they gave participants little opportunity to actively shape the future being presented in the pilot study. As a result, we deliberately integrated opportunities for participant input throughout our probes to prompt reflection and co-speculation.
For example, in LeaseTrack, the contract-aware devices feature input fields to invite participants to customize how these devices should behave or brainstorm and add new devices (Figure ~\ref{fig:linkeddevices}a,c).
Moreover, we designed the probes to have characteristics that are unconventional with `slight strangeness' ~\cite{chow2025beyond, bardzell2013critical} to prompt reflection. 
For example, LeaseCompare speculates an unconventional housing market where contractual clauses are being competed for. On the other hand, contract-aware devices in LeaseTrack aim to provoke reflection on technologies that may overly increase the presence of contractual obligations in daily life.

\subsubsection{Consideration 3: Reducing overlap of design concepts across probes}
During our ideation process, we realized that the same design idea could often be applied to all three contexts (i.e., before, during, and after signing). In our pilot studies, this overlap led to repetitive feedback. 
As a result, we intentionally minimized the repetition of the same design idea across probes. 
For example, while the same concept of Contextualization --- showing relevant legal rights or social comments alongside a clause --- could be applied at any stage, we chose to focus its illustration in LeaseRead when users would be reading and interpreting the contract.
Figure~\ref{fig:featuremapping} provides an overview of the features across the three probes and the design concepts they illustrate.

\subsection{LeaseCompare}
\label{Appendix:LeaseCompare}

\subsubsection{Creating Mock Data}
\label{Appendix:mockdata}

We created mock data for the four apartment listings in LeaseCompare (Figure ~\ref{fig:CompareView}). 
We limited the number of listings to four to prevent overwhelming participants. 
Each listing includes standard details and 14 types of lease clauses. We selected and varied these lease clauses based on two criteria: they are legally significant, and they could vary across leases. To identify them, we reviewed prior legal literature~\cite {leivaditi2020benchmark}, consulted two contract lawyers, examined leases previously signed by the research team, and explored customizable options in online lease creation platforms.\footnote{Online lease creation platforms referenced: LawDepot.com, LawDistrict.com, rocketlawyer.com, Zillow.com, and legaltemplates.net.}

\subsubsection{Calculating and Customizing Composite Scores}
\label{Appendix:compositescore}

Each lease clause is scored based on whether it grants control to the tenant (+1), the landlord (-1), or both parties (+0.5). Then, the scores for clauses within each category (Contract Flexibility, Contract Protection) are summed. The final scores are mapped to letter grades (A-E). The scoring system we used is arbitrary and not meant to prescribe how leases should be ranked. When users rate the relative importance of a clause (Figure ~\ref{fig:LeaseComparecontrol}), the rating (not at all important (0), moderately important (1), or very important (2)) acts as a scaler on how much the clause contributes to the composite score.

\subsection{LeaseRead}
\label{Appendix:LeaseRead}

\subsubsection{Source Document}

The source document of LeaseRead was a lease previously signed by a member of the research team with all identifying information anonymized. We selected 19 sections to augment with one to two features (e.g., information cards, comics, or explorable scenarios) and collapsed the remaining sections, which participants could still expand if desired.

\subsubsection{Guiding Questions}
\label{Appendix:guidingquestions}
When participants engaged with LeaseRead, the following guiding questions were provided to help focus their attention:  

\begin{itemize}
    \item How much do you expect to pay under this contract?
    \item What are some key obligations and dates you need to follow? What happens if you miss them?
    \item What are your and the landlord's rights, responsibilities, or prohibitions?
    \item How is liability distributed between you and your roommate?
    \item What questions would you want to ask the landlord or legal help service before signing?
    \item Other aspects of the lease to consider: landlord entry, replacement/subletting, maintenance, incidents causing personal property loss, and policy changes.
\end{itemize}

\subsubsection{Contextualizing Legal Clauses with External Information}

We used local city ordinances and tenant resources from our university’s legal help office. For resident comments, we referenced tenant comments from Reddit and Apartments.com, and created mock comments about the property and its lease. Figure ~\ref{fig:infocards} shows example Information Cards.

\subsubsection{Comic Scenarios}

To create each comic, we drafted a text scenario illustrating the unintended consequences of a particular section (e.g., a water leak damaged personal property). These text scenarios were input into GPT-4o to generate the initial comic drafts. Since the generated output was not always textually consistent or visually coherent, we refined the comics using Figma.

\subsection{LeaseTrack}
\label{Appendix:LeaseTrack}

The lease content used in LeaseTrack is drawn from the same source contract used in LeaseRead to maintain consistency across the two probes.
Figure~\ref{fig:incidenttracker} displays the full response from Incident Tracker, which we validated with the lead attorney of our university's legal help service. 
Figure~\ref{fig:linkeddevices} shows the comic illustration for the contract-aware speaker. Figure~\ref{fig:linkedglassesandcamera} illustrates the comic illustrations for contract-aware glasses and camera. 
To create these comics, we input the text description of the comics into GPT-4o to obtain the initial comics, which we refined further using Figma.

\subsection{Narrative Scenarios}
\label{subsec:narratives}
We developed narrative scenarios~\cite{chow2025beyond, noortman2019hawkeye} to help participants experience and reflect on each of the three probes. Participants were asked to take on the perspective of Alice, a university student searching for off-campus housing for her second academic year. Alice earns \$3000 per month from a part-time job, owns a pet, does not smoke, and plans to live with a roommate. She is unsure whether she will remain on campus during the summer or renew the lease the following year.

\subsubsection{Choosing an Apartment with LeaseCompare}
In the first scenario, Alice is searching for an apartment using LeaseCompare. Participants were shown four listings with similar amenities but differing rents and lease clauses. To prompt reflection on contractual clauses as a factor in housing decisions, we deliberately assigned higher rents to apartments with more tenant-favorable clauses. Participants were asked to interact with LeaseCompare, review the listings, and select the apartment they would prefer.

\subsubsection{Reading Contract with LeaseRead}
In the second scenario, Alice has identified an apartment and is reviewing the lease before signing using LeaseRead. Guiding questions about the lease content were provided to support participant engagement (Appendix ~\ref{Appendix:guidingquestions}). After reviewing the lease, participants were asked to reflect on the next steps: whether to sign immediately, request clarification, or negotiate specific terms.

\subsubsection{Managing rights and obligations with LeaseTrack}
In the third scenario, Alice has signed the lease and receives a \$250 fee for late rent payment. Participants first reflected on how they would normally respond to such a situation. They then interacted with LeaseTrack to reflect on how the probe might influence their response and obligation management. Additionally, participants browsed through the comic illustrations for each of the three contract-aware devices and were asked to customize device behaviors.

\section{Semi-structured Interview of Past Experiences}
\label{Appendix:interviewquestionsbarriers}

At the start of the study, we conducted a semi-structured interview about participants’ experiences, perceptions, and challenges with leases and other contracts they had previously encountered. The interview was guided by the following questions:

\begin{itemize}
    \item What types of contracts have you interacted with in the past? 
    \item At what points do you typically interact with these contracts?
    \item What resources do you use to help you understand a contract?
    \item What challenges have you encountered when interacting with contracts?
    \item In your view, what are the main purposes of contracts?
    \item To what extent do you think current contracts achieve these purposes?
\end{itemize}

\begin{figure*}[!ht]
    \centering
    \includegraphics[width=0.8\textwidth]{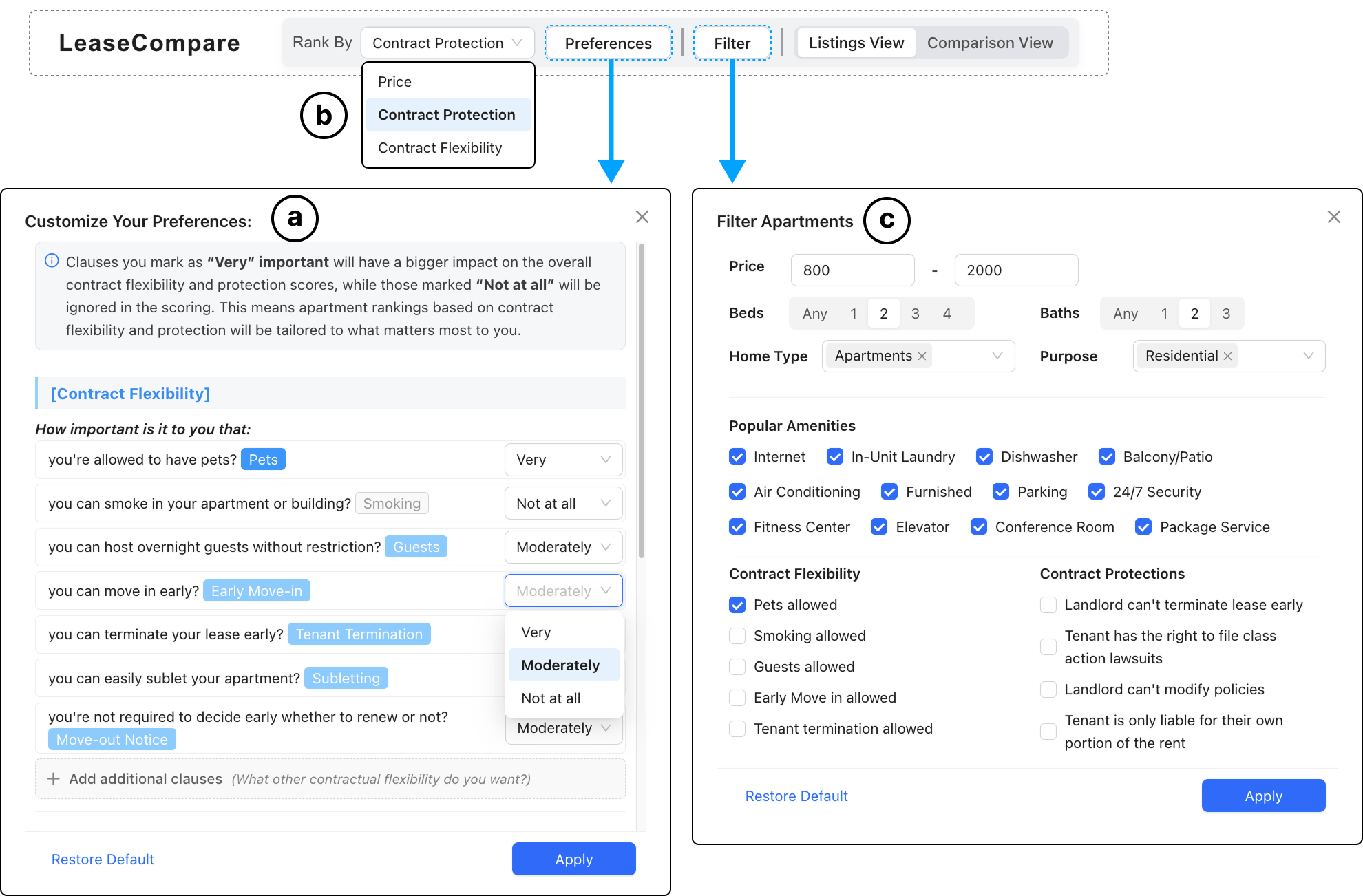}
    \caption{Control Panel in LeaseCompare. Users can (a) adjust preferences on how the two composite scores (i.e., Contract Protection and Flexibility) should be calculated, (b) rank apartments by using the composite scores, and (c) filter apartments by contractual clauses.}
    \label{fig:LeaseComparecontrol}
    \Description{Three pop-up modals displayed when clicking buttons in the top control panel of LeaseCompare. (a) Users can adjust how the two composite scores–-Contract Protection and Contract Flexibility–-are calculated by rating each clause as very, moderately, or not important. (b) Users can select ranking criteria for apartments, choosing among price, Contract Protection Score, or Contract Flexibility Score. (c) Users can set filter criteria to exclude apartments containing certain contractual clauses.}
\end{figure*}

\begin{figure*}[!ht]
    \centering
    \includegraphics[width=0.75\textwidth]{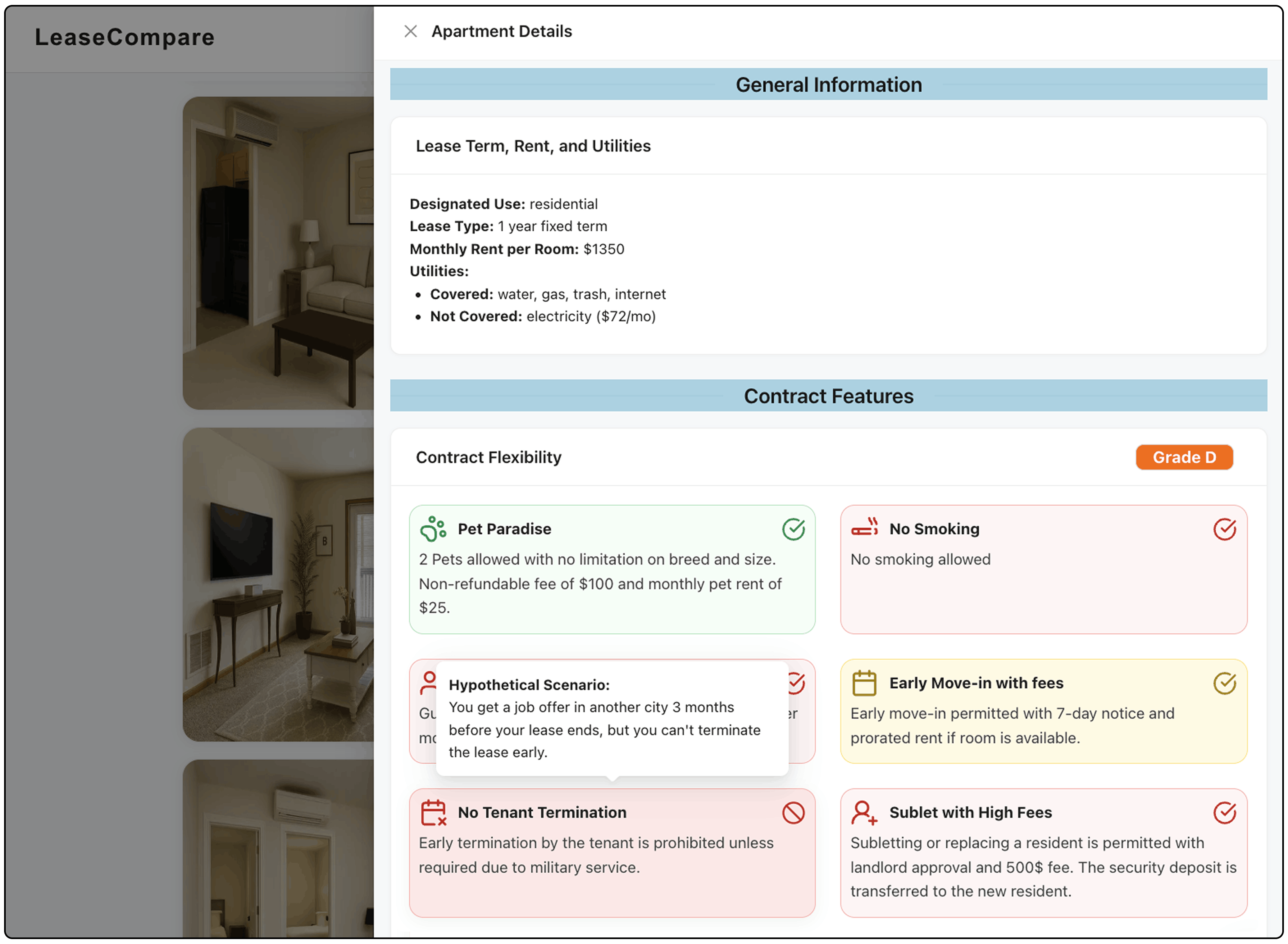}
    \caption{Details View in LeaseCompare. Users can hover over each clause to view a hypothetical scenario.}
    \label{fig:DetailView}
    \Description{Details View in LeaseCompare for an apartment listing. Unlike the Listings View, this view displays the full contractual clauses rather than summaries. Users can hover over each clause to reveal a hypothetical scenario. For example, for a “no tenant termination” clause, the scenario describes receiving a job offer in another city three months before the lease ends but being unable to terminate the lease early.}
\end{figure*}

\begin{figure*}[!ht]
    \centering
    \includegraphics[width=0.95\textwidth]{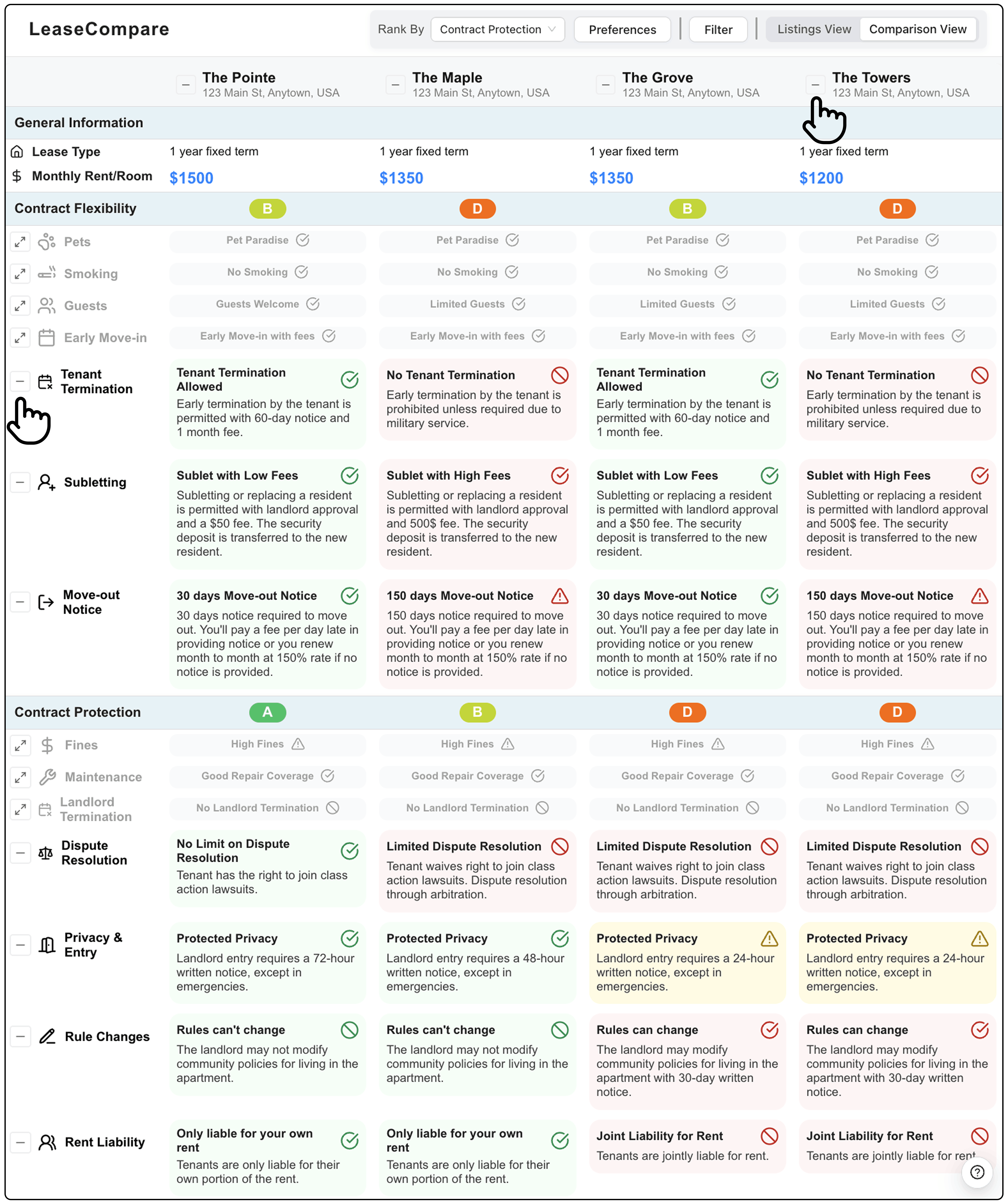}
    \caption{Comparison View in LeaseCompare: An interactive comparison table where users can compare contractual clauses side by side and interactively collapse rows or columns.}
    \label{fig:CompareView}
    \Description{Comparison View in LeaseCompare, showing an interactive table where users can compare contractual clauses side by side and collapse rows or columns. The figure displays four apartments with different prices and lease terms. The Pointe costs \$1,500 and offers the most flexibility (e.g., early move-out allowed) and protection (e.g., right to join a class action lawsuit). The Towers costs \$1,200 and provides the least flexibility (e.g., early move-out prohibited) and protection (e.g., no right to join a class action lawsuit). The Maple and The Grove both cost \$1,350, but differ in contract: The Grove provides more contract flexibility with little protection, while The Maple offers more contract protection with little flexibility.}
\end{figure*}

\begin{figure*}[!ht]
    \centering
    \includegraphics[width=0.95\textwidth]{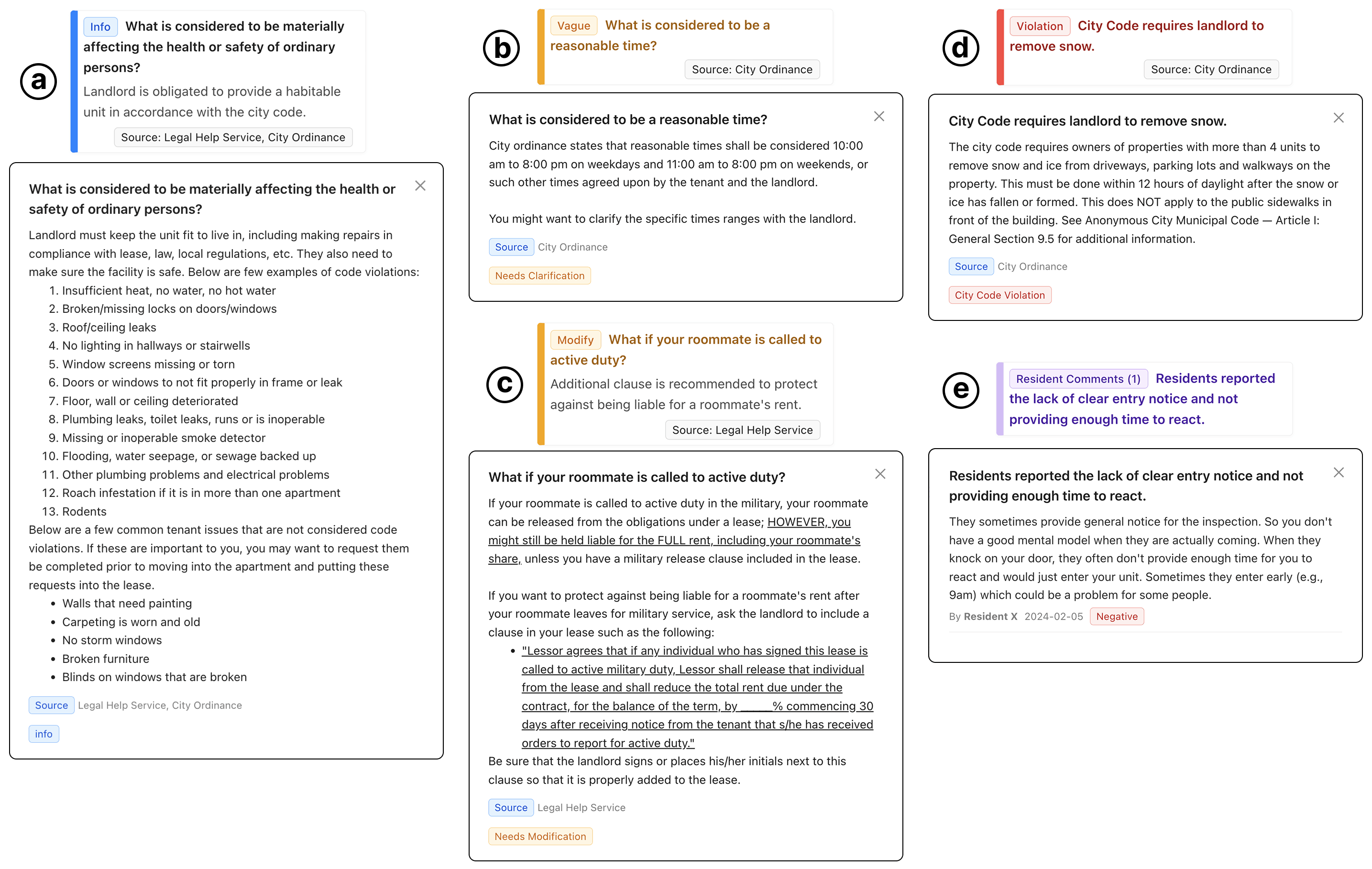}

    \caption{Example Information Cards in LeaseRead: (a) Blue cards explain legal information from local ordinances; (b,c) Yellow cards flag clauses needing clarification or modification; (d) Red cards highlight potentially unlawful clauses; and (e) Purple cards provide social comments.}
    \label{fig:infocards}
    \Description{Five examples of Information Cards in LeaseRead. (a) A blue card explaining the city ordinances about housing conditions that materially affect tenant health or safety, which were not specified in the lease. (b) A yellow card highlighting that a clause did not specify reasonable notice for landlord entry and recommendations from city ordinance to aid negotiation. (c) A yellow card suggesting to add a clause to protect tenants from being liable for a roommate’s rent if the roommate is called to active duty. (d) A red card flagging a conflict where the lease states the landlord is not obligated to remove snow, while ordinances require snow removal. (e) A purple card showing resident comments, such as landlords failing to provide clear entry notice.}
\end{figure*}

\begin{figure*}[!ht]
    \centering
    \includegraphics[width=0.8\textwidth]{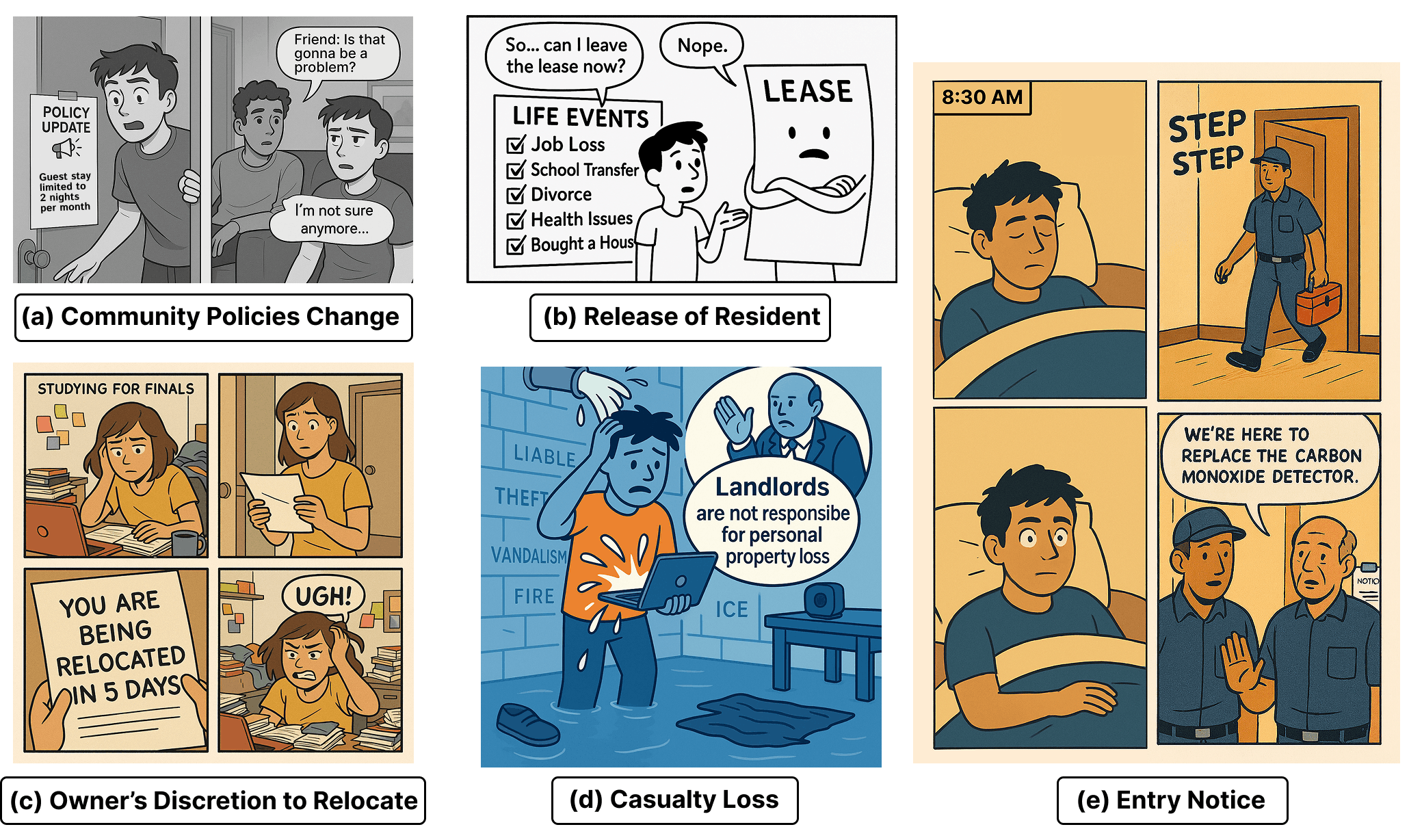}
    \caption{Comic Scenarios in LeaseRead.}
    \label{fig:comicscenarios}
    \Description{The five comic scenarios in LeaseRead. (a) Community Policy Change: The main character invited a friend to stay over, but the next day the landlord posted a new policy limiting guest stays to two days, leaving the tenant uncertain how long the friend could remain. (b) Release of Resident: The tenant asked to terminate the lease early, but the landlord rejected the request. (c) Owner’s Discretion to Relocate: While studying for finals, the tenant received notice that she had to relocate within five days, which left her frustrated. (d) Entry Notice: At 8:30 a.m., the tenant was woken by door knocks as maintenance workers entered the apartment to replace the carbon monoxide detector.}
\end{figure*}

\begin{figure*}[!ht]
    \centering
    \includegraphics[width=0.87\textwidth]{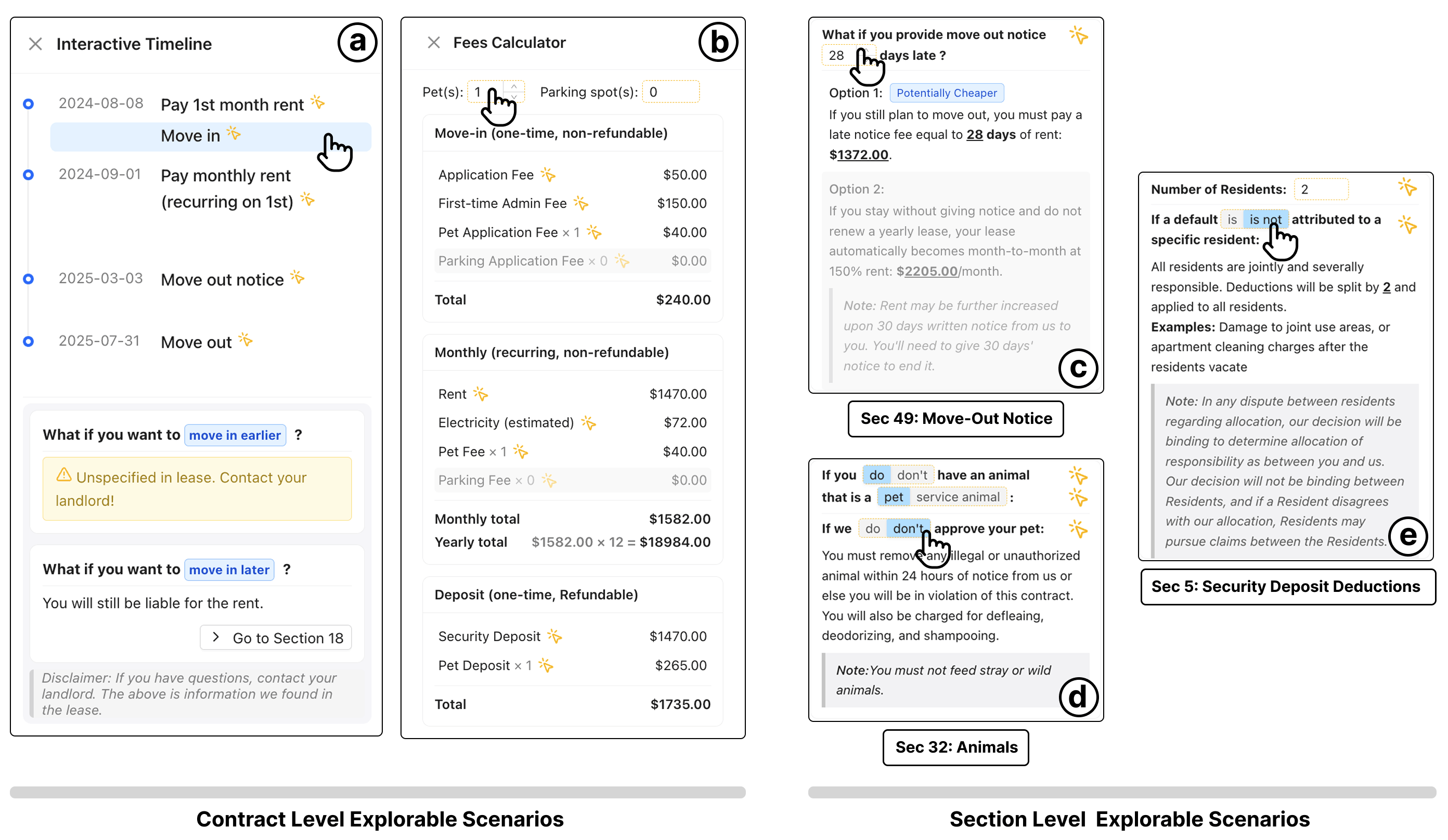}
    \caption{Explorable Scenarios in LeaseRead: (a) Interactive Timeline visualizes time-sensitive obligations and allows users to explore what-if scenarios for being early or late; (b) Fees Calculator aggregates fees and estimates total cost based on pets or parking needed; (c) Cost estimation for providing move-out notice late; (d) Exploration of different situations if having pets or service animals; (e) Exploration of consequences depending on how a default is being attributed.}
    \label{fig:explorablescenarios}
    \Description{The five explorable scenarios in LeaseRead. (a) An interactive timeline visualizing time-sensitive obligations and allows users to click on events to explore what-if scenarios for being early or late. (b) A fees calculator that aggregates fees and estimates total costs based on the number of pets or parking spaces the user need. (c) Users could enter the number of days they were late in providing move-out notice to estimate potential fines. (d) Users can toggle between having pets or service animals to explore relevant lease information. (e) Users can toggle whether a default was attributed to all residents versus specific residents to view the resulting fines.}
\end{figure*}

\begin{figure*}[!ht]
    \centering
    \includegraphics[width=0.83\textwidth]{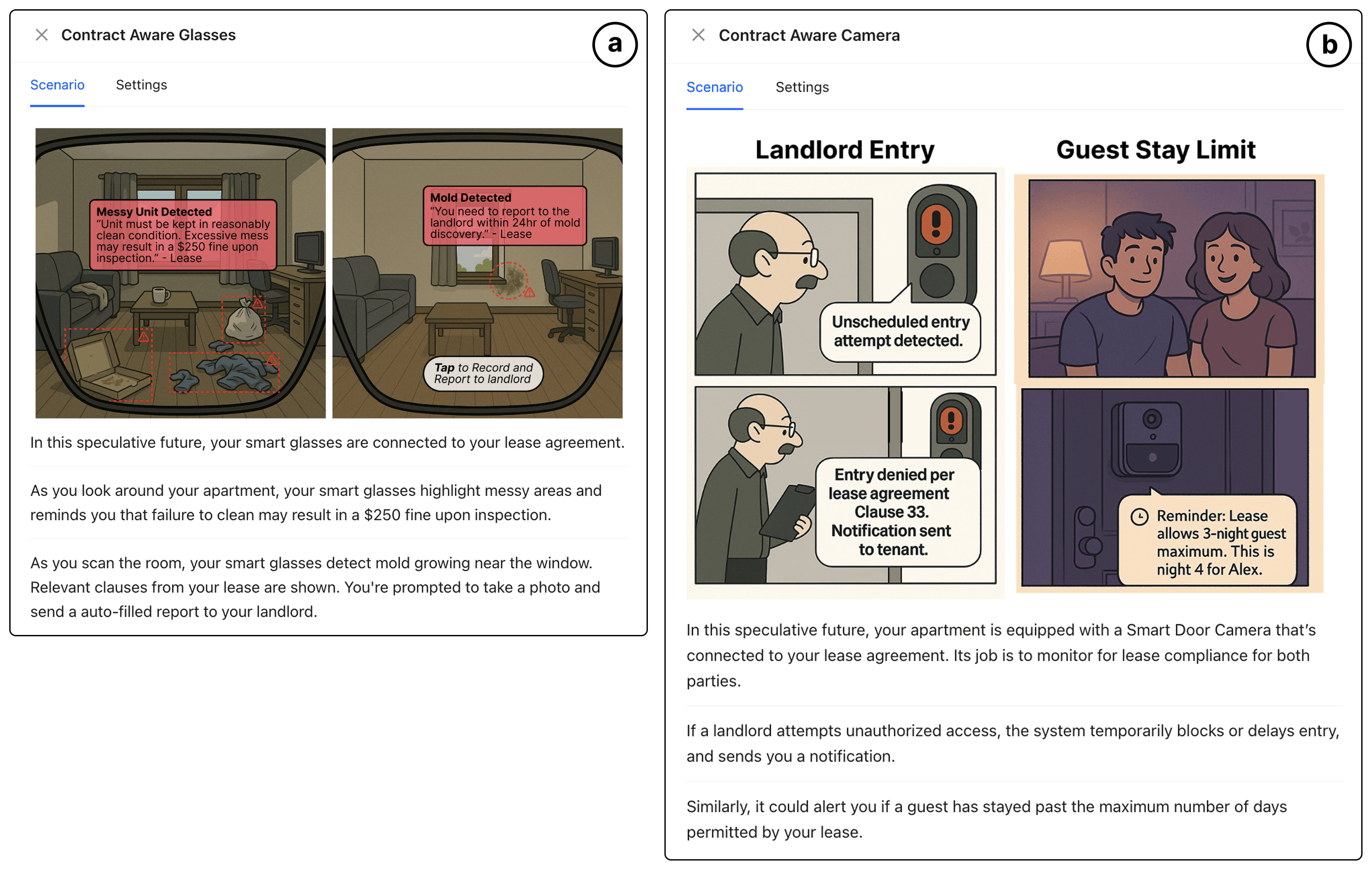}
    \caption{Comic illustrations and narratives for (a) Contract Aware Glasses and (b) Contract Aware Camera in LeaseTrack.}
    \label{fig:linkedglassesandcamera}
    \Description{Comic illustrations and narratives for (a) Contract Aware Glasses and (b) Contract Aware Camera in LeaseTrack. Contract Aware Glasses: `In this speculative future, your smart glasses are connected to your lease agreement. As you look around your apartment, your smart glasses highlight messy areas and reminds you that failure to clean may result in a \$250 fine upon inspection. As you scan the room, your smart glasses detect mold growing near the window. Relevant clauses from your lease are shown. You're prompted to take a photo and send a auto-filled report to your landlord.’ Contract Aware Camera: `In this speculative future, your apartment is equipped with a Smart Door Camera that’s connected to your lease agreement. Its job is to monitor for lease compliance for both parties. If a landlord attempts unauthorized access, the system temporarily blocks or delays entry, and sends you a notification. Similarly, it could alert you if a guest has stayed past the maximum number of days permitted by your lease.’}
\end{figure*}

\begin{figure*}[!ht]
    \centering
    \includegraphics[width=1\textwidth]{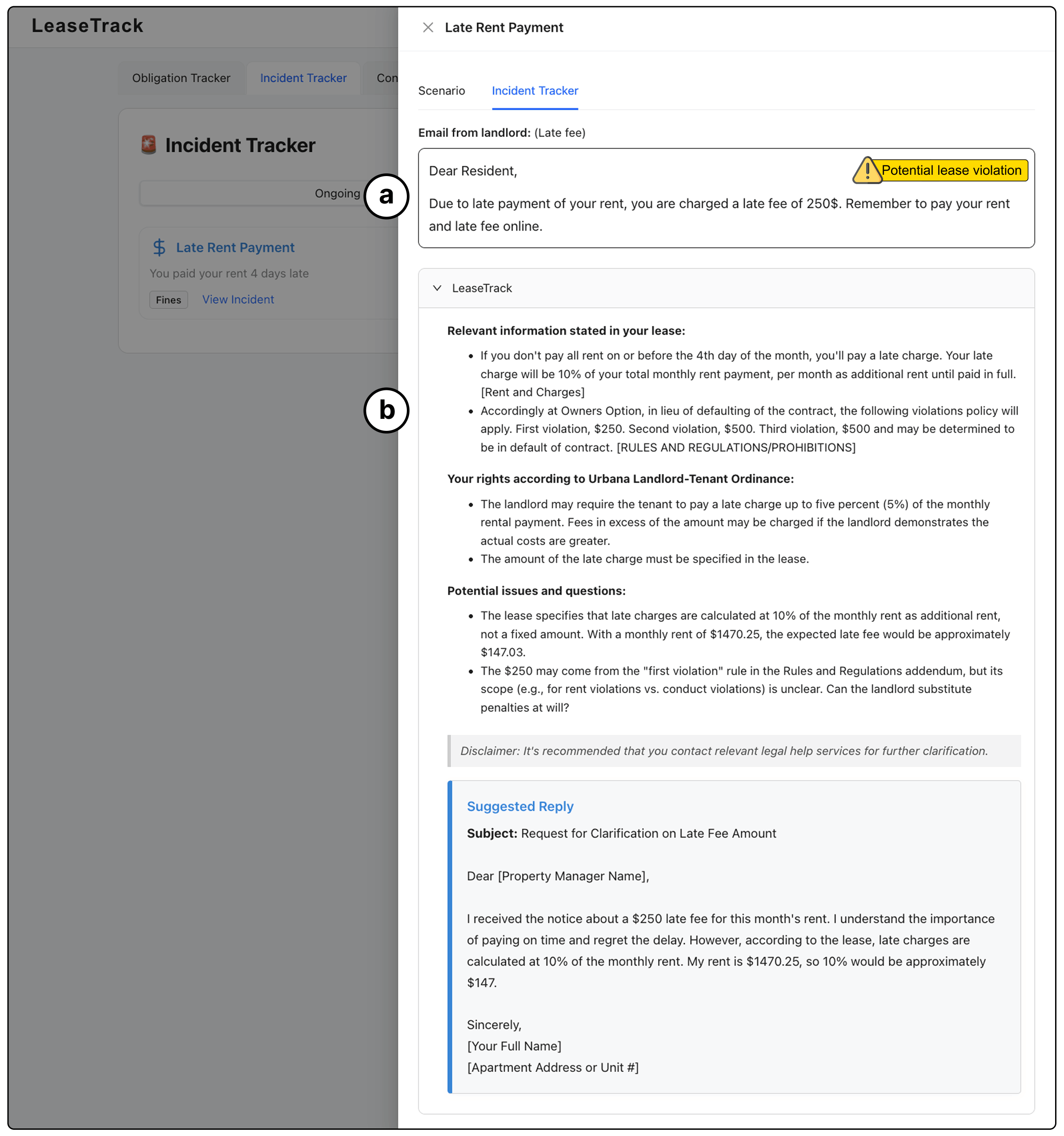}
    \caption{Incident Tracker in LeaseTrack: (a) A hypothetical scenario of a landlord trying to overcharge a late fee through an email, which the Incident Tracker flags as a potential violation of the lease; (b) Incident Tracker provides relevant information from the lease and local ordinance, along with a suggested email reply.}
    \label{fig:incidenttracker}
    \Description{Incident Tracker in LeaseTrack showing two panels. (a) A hypothetical email from the landlord trying to overcharge late fee `Dear Resident, Due to late payment of your rent, you are charged a late fee of 250\$. Remember to pay your rent and late fee online.’; (b) Incident Tracker flags the email as a potential violation of the lease, supply relevant information from the lease and local ordinance, and suggests an email reply. `Relevant information stated in your lease: If you don't pay all rent on or before the 4th day of the month, you'll pay a late charge. Your late charge will be 10\% of your total monthly rent payment, per month as additional rent until paid in full; Your rights according to Urbana Landlord-Tenant Ordinance: The amount of the late charge must be specified in the lease; Suggested Reply: I received the notice about a \$250 late fee for this month's rent. I understand the importance of paying on time and regret the delay. However, according to the lease, late charges are calculated at 10\% of the monthly rent. My rent is \$1470.25, so 10\% would be approximately \$147.’}
\end{figure*}

\aptLtoX[graphic=no,type=html]{\begin{table*}[hb]
\centering
\renewcommand{\arraystretch}{1.2} 
\setlength{\tabcolsep}{5pt} 
\begin{tabular}{p{2cm} p{2.5cm} p{3cm} p{8cm}}
\hline
\textbf{Feature} & 
\textbf{Design Concepts} & 
\textbf{Target Contracts} & 
\textbf{Example Feature Description} \\
\hline

\textcolor{black}{\textbf{\cellcolor[HTML]{e5e5e5}{Before Signing}}}&\cellcolor[HTML]{e5e5e5}&\cellcolor[HTML]{e5e5e5}&\cellcolor[HTML]{e5e5e5}\\ 

\hline

Compare Contracts (10) &  
Representation, Proactivity & 
Mortgage, Loan, Rental, Insurance, ToS

& 
\textit{``I think the comparison view would be very helpful when it comes to insurance, especially health. Allows you to compare the different ways in which you are covered and not covered, and the copays.''} (P3, Insurance) \\
\hline

\textcolor{black}{\textbf{\cellcolor[HTML]{e5e5e5}{While Signing}}}&\cellcolor[HTML]{e5e5e5}&\cellcolor[HTML]{e5e5e5}&\cellcolor[HTML]{e5e5e5}\\ 

\hline

Information Cards (15) &  
Contextualization & 
Mortgage, Loan, Rental, Insurance, Surgical, Employment, ToS & 
\textit{``The system could tell you the law in your State. Such as if I'm hurt on the job, who is responsible for my medical care? Also, I think anything [in the contract] having to do with possibly breaking the law would be important to know.''} (P17, Employment) \\
\hline

Explorable Scenarios (15) &  
Representation & 
Mortgage, Loan, Rental, Insurance, Surgical, Employment, ToS & 
\textit{``I would love to see the fee calculator with each detailed fee. So I can explore if I do a different type of installment plan, early payoff, deferment, or refinance, what would that look like altogether? ''} (P10, Student Loan) \\
\hline

Multi-Contract Integration (1) &  
Representation & 
Insurance, Surgical & 
\textit{``For interactive scenarios, in addition to calculating the fee based on the number of nights [in the hospital], it allows you to see whether you're covered or not...''} (P7, Surgical and Insurance) \\
\hline

Comic Scenarios (2) &  
Representation & 
Insurance, ToS & 
\textit{``for example, I think especially when I think back to the comics that was on showing someone like, Oh, you ran out of gas. You're on the highway, or your tire went flat. A lot of companies. They have the roadside assistance.
''} (P2, Car Insurance) \\
\hline

Interactive Quiz (1) &  
Representation & 
Rental & 
\textit{``Before you sign the rental contract, it can give you a quiz. It requires you to pass a quiz to make sure that you understand the contract before you sign it.''} (P16, Rental) \\
\hline

Searchable Contract (1) &  
Representation & 
Employment & 
\textit{`` I'd want a better search feature than all the email search features, to actually find relevant information...''} (P17, Employment) \\
\hline

\textcolor{black}{\textbf{\cellcolor[HTML]{e5e5e5}{After Signing}}}&\cellcolor[HTML]{e5e5e5}&\cellcolor[HTML]{e5e5e5}&\cellcolor[HTML]{e5e5e5}\\ 
\hline

Obligation Tracker (8) &  
Representation, Proactivity & 
Mortgage, Loan, Rental, Insurance, Surgical, Employment, ToS & 
\textit{``The Obligation Tracker could give you a visual of your process and progress with paying off your loans. Like, how much have I paid so far?
When did I stop or get a deferment? 
When did I not pay any interest on it.''} (P10, Student Loan)  \\
\hline

Incident Tracker (6) &  
Proactivity & 
Insurance, Rental, Surgical, Employment, ToS & 

\textit{``Maybe at the moment, you're getting the MRI done. You don't really have the energy to read the contract. Later, go to the Incident Tracker and then find out what my rights are. Maybe I only have 90 days after to appeal.''} (P10, Medical Insurance)  \\
\hline

Contract Aware Devices (8) &  
Proactivity & 
Rental, Insurance, Surgical, Employment & 
\textit{``It could have a front or back camera on the car recording for insurance purposes for car rentals. If you're in an accident, you do panic in that situation. The contract awareness devices could help you with starting a claim.''} (P8, Car Rental and Insurance)  \\
\hline

Track Contract Changes (1) &  
Proactivity & 
ToS & 
\textit{`` When the contract changes, I want an AI to tell me, they changed line 7 on page 17, under subparagraph 4. And now they're allowed to take your firstborn if you're late on a payment. ''} (P4, Credit Card ToS)  \\

\hline
\end{tabular}
\caption{Examples of how participants adopted the design probes to contracts other than leases in the design ideation session (RQ3). For ideas resembling features of the probes (e.g., Information Cards, Explorable Scenarios, etc.), we used their corresponding names.}
\label{table:participantdesignideas}
\end{table*}}{\begin{table*}[hb]
\centering
\renewcommand{\arraystretch}{1.2} 
\setlength{\tabcolsep}{5pt} 
\begin{tabular}{p{2cm} p{2.5cm} p{3cm} p{8cm}}
\hline
\textbf{Feature} & 
\textbf{Design Concepts} & 
\textbf{Target Contracts} & 
\textbf{Example Feature Description} \\
\hline

\rowcolor{gray!20} 
\multicolumn{4}{l}{\textbf{Before Signing}} \\ 
\hline

Compare Contracts (10) &  
Representation, Proactivity & 
Mortgage, Loan, Rental, Insurance, ToS

& 
\textit{``I think the comparison view would be very helpful when it comes to insurance, especially health. Allows you to compare the different ways in which you are covered and not covered, and the copays.''} (P3, Insurance) \\
\hline

\rowcolor{gray!20} 
\multicolumn{4}{l}{\textbf{While Signing}} \\ 
\hline

Information Cards (15) &  
Contextualization & 
Mortgage, Loan, Rental, Insurance, Surgical, Employment, ToS & 
\textit{``The system could tell you the law in your State. Such as if I'm hurt on the job, who is responsible for my medical care? Also, I think anything [in the contract] having to do with possibly breaking the law would be important to know.''} (P17, Employment) \\
\hline

Explorable Scenarios (15) &  
Representation & 
Mortgage, Loan, Rental, Insurance, Surgical, Employment, ToS & 
\textit{``I would love to see the fee calculator with each detailed fee. So I can explore if I do a different type of installment plan, early payoff, deferment, or refinance, what would that look like altogether?''} (P10, Student Loan) \\
\hline

Multi-Contract Integration (1) &  
Representation & 
Insurance, Surgical & 
\textit{``For interactive scenarios, in addition to calculating the fee based on the number of nights [in the hospital], it allows you to see whether you're covered or not...''} (P7, Surgical and Insurance) \\
\hline

Comic Scenarios (2) &  
Representation & 
Insurance, ToS & 
\textit{``for example, I think especially when I think back to the comics that was on showing someone like, Oh, you ran out of gas. You're on the highway, or your tire went flat. A lot of companies. They have the roadside assistance.''} (P2, Car Insurance) \\
\hline

Interactive Quiz (1) &  
Representation & 
Rental & 
\textit{``Before you sign the rental contract, it can give you a quiz. It requires you to pass a quiz to make sure that you understand the contract before you sign it.''} (P16, Rental) \\
\hline

Searchable Contract (1) &  
Representation & 
Employment & 
\textit{`` I'd want a better search feature than all the email search features, to actually find relevant information...''} (P17, Employment) \\
\hline

\rowcolor{gray!20} 
\multicolumn{4}{l}{\textbf{After Signing}} \\ 
\hline

Obligation Tracker (8) &  
Representation, Proactivity & 
Mortgage, Loan, Rental, Insurance, Surgical, Employment, ToS & 
\textit{``The Obligation Tracker could give you a visual of your process and progress with paying off your loans. Like, how much have I paid so far?
When did I stop or get a deferment? 
When did I not pay any interest on it.''} (P10, Student Loan)  \\
\hline

Incident Tracker (6) &  
Proactivity & 
Insurance, Rental, Surgical, Employment, ToS & 

\textit{``Maybe at the moment, you're getting the MRI done. You don't really have the energy to read the contract. Later, go to the Incident Tracker and then find out what my rights are. Maybe I only have 90 days after to appeal.''} (P10, Medical Insurance)  \\
\hline

Contract Aware Devices (8) &  
Proactivity & 
Rental, Insurance, Surgical, Employment & 
\textit{``It could have a front or back camera on the car recording for insurance purposes for car rentals. If you're in an accident, you do panic in that situation. The contract awareness devices could help you with starting a claim.''} (P8, Car Rental and Insurance)  \\
\hline

Track Contract Changes (1) &  
Proactivity & 
ToS & 
\textit{`` When the contract changes, I want an AI to tell me, they changed line 7 on page 17, under subparagraph 4. And now they're allowed to take your firstborn if you're late on a payment.''} (P4, Credit Card ToS)  \\

\hline
\end{tabular}
\caption{Examples of how participants adopted the design probes to contracts other than leases in the design ideation session (RQ3). For ideas resembling features of the probes (e.g., Information Cards, Explorable Scenarios, etc.), we used their corresponding names.}
\label{table:participantdesignideas}
\Description{In the design session, participants extended the probes to contracts beyond leases. The table lists example ideas they proposed. For ideas resembling probe features (e.g., Information Cards, Explorable Scenarios), we used their original names. For each feature, the table shows its corresponding Living Contracts design concept (i.e., Contextualization, Representation, or Proactivity), the types of contracts participants applied it to, and an example quote describing the feature.}
\end{table*}}

\begin{table*}[!htb]
\centering
\setlength{\tabcolsep}{3pt} 
\renewcommand{\arraystretch}{1.05} 
\caption{Participant Demographics}
\label{tab:demographics}
\Description{Participant demographics: All participants were fluent in English, and one participant self-identified as a non-native speaker (P5). One participant was a landlord with experience reviewing and drafting leases (P4), four described regularly interacting with contracts or legal documents as part of their work (P1, P3, P10, P14), and one had completed graduate-level coursework related to contract law (P12).}
\begin{tabular}{lllllll}
\toprule
\textbf{PID} & \textbf{Age} & \textbf{Sex} & \textbf{Ethnicity} & \textbf{Education} & \textbf{Occupation} \\
\midrule
P1  & 63 & Female & White & Associate's in Economics & Mortgage Banker \\
P2  & 36 & Female & Black & Bachelor's in Business & Small Business Owner \\
P3  & 54 & Female & White & Bachelor's in Communication & Legal Scopist  \\
P4  & 48 & Male   & White & Associate's in Accounting & Landlord, Process Manager  \\
P5  & 44 & Female & White & Master's in Engineering & Director of Engineering \\
P6  & 26 & Female & White & Associate's in Science & Commission Artist  \\
P7  & 49 & Female & White & PhD in Psychology & Consultant  \\
P8  & 49 & Female & White & High School & Freelancer  \\
P9  & 37 & Male   & Mixed & Associate's in Art and Design & Small Business Owner  \\
P10 & 38 & Female & Black & Associate's in Cybersecurity & Loan Officer  \\
P11 & 42 & Female & White & Bachelor's in Behavior Health & Manufacturing Technician  \\
P12 & 37 & Male   & Mixed & Master's in Business and Legal Administration &  Consultant \\
P13 & 34 & Male   & Black & Bachelor's in Education & Security  \\
P14 & 63 & Female & White & High School & Regulatory Compliance Specialist  \\
P15 & 36 & Male   & White & Bachelor's in Economics & ESL Teacher  \\
P16 & 30 & Male   & Asian & Master's in Accounting & Accountant  \\
P17 & 48 & Female & White & Associate's in Communication & Non-profit Program Manager  \\
P18 & 46 & Male   & White & PhD in Information Technology & Student  \\
\bottomrule
\end{tabular}
\end{table*}


\end{document}